# Electrohydrodynamic migration of a spherical drop in a general quadratic flow

Shubhadeep Mandal[1], Aditya Bandopadhyay[2] and Suman Chakraborty[1,2]†

[1]Department of Mechanical Engineering, Indian Institute of Technology Kharagpur, West Bengal- 721302, India

[2]Advanced Technology Development Center, Indian Institute of Technology Kharagpur, West Bengal- 721302, India

We investigate the motion of a spherical drop in a general quadratic flow acted upon by an arbitrarily oriented externally applied uniform electric field. The drop and media are considered to be leaky dielectrics. The flow field affects the distribution of charges on the drop surface, which leads to alteration in the electric field, thereby affecting the velocity field through the Maxwell stress on the fluid-fluid interface. The two-way coupled electrohydrodynamics is central towards dictating the motion of the drop in the flow field. We analytically address the electric potential distribution and Stokesian flow field in and around the drop in a general quadratic flow for small electric Reynolds number (which is the ratio of the charge relaxation time scale to the convective time scale). As a special case, we consider a drop in an unbounded cylindrical Poiseuille flow and show that, an otherwise absent, cross-stream drop migration may be obtained in the presence of a uniform electric field. Depending on the direction of the applied electric field and the relative electrical properties of the drop and media, the drop may migrate towards or away from the centreline of the imposed Poiseuille flow.

_______________________________________________________________

# 1. Introduction

Drops play a central role in interdisciplinary microfluidic and nanofluidic research (Casadevall i Solvas & DeMello 2011; Seemann et al. 2012; Teh et al. 2008). They form a convenient means for rapid analytic detection and screening of chemicals (Zheng et al. 2004; Zhu & Fang 2013), protein crystallization (Zhu et al. 2014), medium for cell/particle encapsulation (Bhagat et al. 2010), reagent mixing (Bringer et al. 2004), and biological assays (Guo et al. 2012). Precise control and manipulation of drops is of utmost importance for realizing optimal functionalities in these concerned applications. Towards this, the effective manipulation of drops by various means such as thermocapillary (Baroud et al. 2007; Basu & Gianchandani 2008), acoustic streaming (Franke et al. 2009), electric fields (Link et al. 2006; Ahn et al. 2006), magnetic fields (Pamme 2012) etc. have been discussed by various researchers. The above processes typically involve low Reynolds numbers based on the drop size and velocity. The cross-stream migration of a non-deforming spherical drop with clean fluid-fluid interface, solely due to such a creeping flow field, is not possible, since the governing equations and boundary conditions are linear in nature (Leal 2007; Hanna & Vlahovska 2010). The various mechanisms which lead to a cross-stream migration of the drop include drop deformation (Haber & Hetsroni 1971; Griggs et al. 2007; Wohl & Rubinow 2006; Chan & Leal 2006), non-negligible fluid inertia (Mortazavi & Tryggvason 2000; Magnaudet 2003), flow induced surfactant redistribution on the drop surface (Hanna & Vlahovska 2010; Pak et al. 2014), fluid viscoelasticity (Chan & Leal 2006; Mukherjee & Sarkar 2013; Mukherjee & Sarkar 2014) etc. The underpinning non-linearity in either the boundary conditions or the governing equations is the common tie between the aforementioned mechanisms causing cross-stream migration.

†Email address for correspondence: suman@mech.iitkgp.ernet.in

The ease of integration and flexibility of operation renders external electric fields as a convenient means for drop manipulation in modern microfluidic devices. Studies on response of drops towards externally applied electric field have been carried out since the seminal work of Taylor (1966). Taylor (1966) introduced the leaky dielectric model which considers small electrical conductivity of liquids and disregarded the effect of charge convection. Several researchers have employed this model and studied the deformation of neutrally buoyant drops in the presence of uniform or non-uniform electric fields in an otherwise quiescent medium (Vizika & Saville 2006; Supeene et al. 2008; Lanauze et al. 2013; Deshmukh & Thaokar 2013; Torza et al. 1971; Lac & Homsy 2007; Feng & Scott 2006; Feng 1999; Salipante & Vlahovska 2010; Ward & Homsy 2006; Ward & Homsy 2001; Xu & Homsy 2007). Most of these studies considered one way coupled electrohydrodynamics - the applied electric field affects the flow filed via generation of Maxwell stress at the fluid-fluid interface while the electric field remains unaltered by the flow field. However, there appears to be scanty literature which considers the coupled effect of electric field and background flow field towards dictating the motion and deformation of drops (Xu & Homsy 2006; Vlahovska 2011; He et al. 2013; Schwalbe et al. 2011). Presence of a background flow field has shown to markedly alter the charge distribution on the drop surface which further alters the drop motion and deformation. Xu and Homsy (2006) considered the sedimentation of a non-neutrally buoyant drop in the presence of an axial uniform electric field. They concluded that the effect of charge convection has a profound impact on the settling velocity of the drop. In a later study, Vlahovska (2011) performed a perturbation analysis to study the drop deformation, orientation characteristics and shear rheology of highly viscous drops in a uniform electric field in the presence of background linear flows. The background flow induces an asymmetry in the surface charge distribution, thereby fundamentally altering the electric field distribution in and around the drop.

In many practical scenarios, suspending drops are transported by the application of a pressure gradient using syringe pumps. In such situations, it is quite expected that the drops encounter quadratic flow fields. Motivated by this consideration, here we analyze the non-trivial implication of the curvature of the background quadratic flow field in the presence of a uniform electric field. Notably, the asymmetry brought about by the flow through the surface charge convection is expected to be even more prominent in the case of quadratic flows as compared to linear flows, more so in the case when the electric field is non-aligned with the general flow direction. Thus, going beyond the considerations of a linear flow, we analyze the motion of a Newtonian leaky dielectric drop in a general quadratic flow (and present specific results for cylindrical Poiseuille flow) of another Newtonian leaky dielectric medium. The applied electric field is considered to be uniform but oriented in an arbitrary direction. In order to isolate the effect of surface charge convection, and for the sake of analytical tractability, we consider a non-deforming spherical drop in Stokes flow. This assumption is valid for limitingly small value of capillary number which is a manifestation of very strong surface tension force (Leal 2007; Pak et al. 2014; Hanna & Vlahovska 2010; Subramanian & Balasubramaniam 2005). Considering weak surface charge convection, as manifested through the perturbation parameter $Re_E$ (the electric Reynolds number - it signifies the ratio of charge relaxation timescale to the convective time scale), we arrive at general expressions for the drop velocity and the flow field inside and outside the drop through a coupling of the background fluid flow and electric field. Our major finding in the present study is that the simultaneous application of an electric field in the longitudinal and transverse direction yields a cross-stream migration which is dependent on the relative ratios of fluid properties such as permittivity, conductivity and viscosity. Our results also indicate that depending on the relative magnitudes of the ratios of permittivity and conductivity, the drop migration in cylindrical Poiseuille flow may be towards or away from the centreline of the flow. The work aims at unveiling the prospect of indirect drop control in generalized flows via the coupled influence of the flow field and electric field. The methodology outlined here is expected to yield better drop control over other dielectrophoretic methods which employ non-uniform electric fields or thermocapillarity based methods where rapid switching and fast response times are difficult to obtain.

## 2. Problem formulation

We consider the motion of a spherical drop of radius $a$ in an unbounded domain in the presence of an imposed quadratic flow field $\left( \mathbf{V}_{\infty} \right)$ and a uniform electric field $\left( \mathbf{E}_{\infty} \right)$. Here we consider the general quadratic flow field of the form $\mathbf{V}_{\infty} = V_c \left( k_0 + k_1 x + k_2 x^2 + k_3 y^2 \right) \mathbf{e}_z$, and the imposed electric field of the form $\mathbf{E}_{\infty} = E_c \left( E_x \mathbf{e}_x + E_y \mathbf{e}_y + E_z \mathbf{e}_z \right)$ with $V_c$ and $E_c$ being the characteristic velocity and characteristic magnitude of the electric field respectively. Different kinds of imposed flow configurations can be obtained by properly specifying the values of $k_0, k_1, k_2$ and $k_3$ while the direction of the applied electric field can be altered by properly specifying the components of the imposed electric field $\left( E_x, E_y \text{ and } E_z \right)$. The drop liquid is considered to be Newtonian with a viscosity $\lambda_i$, density $\rho_i$, and leaky dielectric with an electrical conductivity $\sigma_i$ and permittivity $\varepsilon_i$. The suspending fluid is also considered as Newtonian and leaky dielectric with the hydrodynamic and electrical properties represented by the same symbols, but using the subscript $e$. The spherical coordinate system $\left(r, \theta, \phi\right)$ moves with the drop at a velocity $\mathbf{U}_d$ which is yet to be determined (please refer to figure 1). Here, we assume that the capillary number is characteristically low to cause any finite deformation of the drop (Hanna & Vlahovska 2010; Pak et al. 2014). We also assume that the Reynolds number based on drop radius is very small so that the fluid inertia is completely negligible (Pak et al. 2014; Hanna & Vlahovska 2010; Vlahovska et al. 2005). We consider the drop as neutrally buoyant $\left(\text{i.e. } \rho_i = \rho_e\right)$ so that gravity has no effect on the motion of drop.

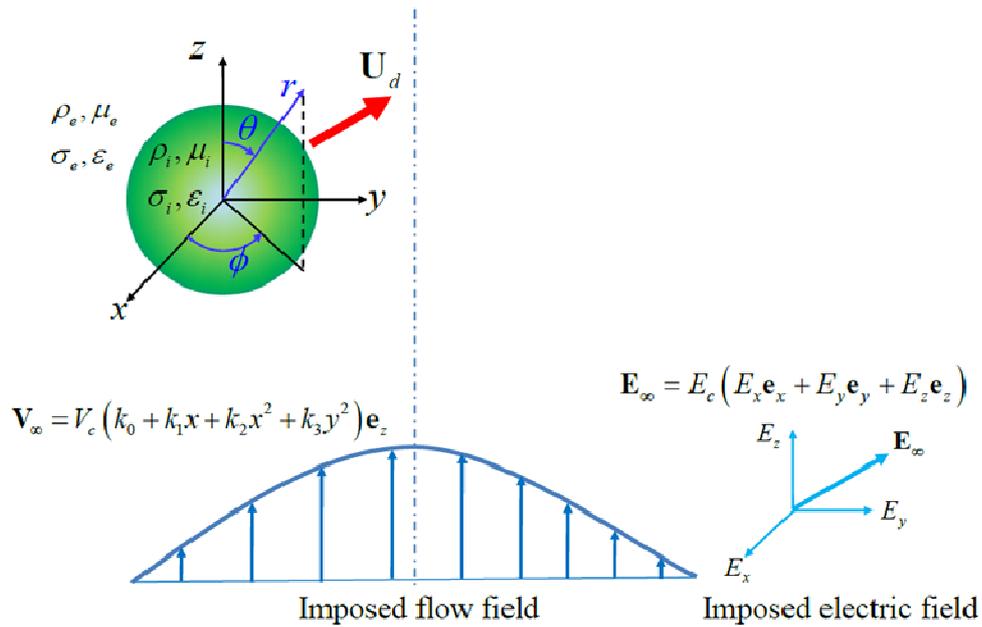

FIGURE 1. (Colour online) Schematic of a spherical drop in a general quadratic flow field $\mathbf{V}_\infty$ in the presence of an imposed electric field $\mathbf{E}_\infty$ in an arbitrary direction. A spherical coordinate system $\left(r, \theta, \phi\right)$ is considered which is attached to the drop centre. The density, viscosity, permittivity and conductivity of the drop liquid are denoted by $\rho_i, \mu_i, \varepsilon_i$ and $\sigma_i$ respectively; while the same symbols with subscript $e$ denote the hydrodynamic and electrical properties of the external liquid.

We non-dimensionalize the length by the radius of the spherical drop $a$. The velocity scale is taken as $V_c$ which is the characteristic velocity of the imposed quadratic flow $\mathbf{V}_\infty$. The characteristic viscous stress and electrical stresses are taken as $\tau_c^H = \mu_e V_c / a$ and $\tau_c^E = \varepsilon_e E_c^2$ respectively. Here we consider the properties of the external liquid as the characteristic quantities which yields the following property ratios (Xu & Homsy 2006): viscosity ratio $\lambda = \mu_i / \mu_e$, conductivity ratio $R = \sigma_i / \sigma_e$ and permittivity ratio $S = \varepsilon_i / \varepsilon_e$. Using this non-dimensional scheme we obtain the following important non-dimensional numbers: electric Reynolds number $Re_E = \varepsilon_e V_c / a \sigma_e$ and Mason parameter $M = a \varepsilon_e E_c^2 / \mu_e V_c$.

The electric Reynolds number signifies the relative importance of the charge relaxation time scale as compared to the charge convection time scale, whereas the Mason parameter signifies the relative importance of the electrical stresses as compared to the viscous stresses.

## 2.1. *Governing equations and boundary conditions*

The drop and the suspending medium are considered as weakly conducting liquids under the paradigm of the leaky dielectric model (Taylor 1966; Saville 1997; Melcher & Taylor 1969). The leaky dielectric model assumes the bulk liquids as charge-free, and the mismatch between the conductivity and permittivity of the two liquids manifested in terms of the interfacial charge at the drop surface only. Hence, the electric potential $\left(\psi_{i,e}\right)$ satisfies the Laplace equation inside and outside the drop

$$\nabla^2 \psi_{i,e} = 0. \tag{2.1}$$

The electric potential inside and outside the drop satisfy the following boundary conditions (Xu & Homsy 2006; Vlahovska 2011):

(e1) the electric potential inside the drop $\psi_i$ is bounded at the origin of the spherical coordinate system,

(e2) the electric potential outside the drop $\psi_e$ approaches the specified potential at infinity $\psi_\infty$, which is given in terms of electric potential as: $\nabla \psi_\infty = -\mathbf{E}_\infty$,

(e3) the electric potential is continuous at the surface of the drop: at $r = 1$, $\psi_i = \psi_e$,

(e4) the charges present at the drop surface satisfy the following conservation equation: at $r = 1$, $\mathbf{e}_r \cdot \left( R \nabla \psi_i - \nabla \psi_e \right) = -Re_E \nabla_s \cdot \left( q_s \mathbf{V}_s \right)$, where the surface charge is given by $q_s = \mathbf{e}_r \cdot \left( S \nabla \psi_i - \nabla \psi_e \right)$, $\mathbf{V}_s$ is the fluid velocity at the drop surface and $\nabla_s = \left[ \nabla - \mathbf{e}_r \left( \mathbf{e}_r \cdot \nabla \right) \right]$ represents the surface divergence operator.

The velocity field inside and outside the drop satisfy the Stokes equation and the condition of incompressibility of the form (Happel & Brenner 1981):

$$\nabla p_{i,e} = N_\mu \nabla^2 \mathbf{u}_{i,e}, \ \nabla \cdot \mathbf{u}_{i,e} = 0, \tag{2.2}$$

where $\left( \mathbf{u}_i, p_i \right)$ and $\left( \mathbf{u}_e, p_e \right)$ represent the velocity and the pressure fields inside and outside the drop, respectively. The term $N_\mu$ is the normalised viscosity; $N_\mu = \lambda$ inside the drop and $N_\mu = 1$ outside the drop.

The flow field inside and outside the drop satisfy the following boundary conditions:

(f1) the velocity $\left(\mathbf{u}_i\right)$ and pressure $\left(p_i\right)$ inside the drop are bounded at the origin of the spherical coordinate system,

(f2) the velocity outside the drop $\mathbf{u}_e$ approaches the specified imposed quadratic velocity at infinity, which in a reference frame attached to the drop centre, can be given by: $\mathbf{u}_e\big|_{r\to\infty} = \mathbf{u}_\infty = \mathbf{V}_\infty - \mathbf{U}_d$,

(f3) the velocity is continuous at the drop surface and at steady state the normal component of the velocity at the drop surface is zero; at $r = 1$, $\mathbf{u}_i = \mathbf{u}_e$ and $\mathbf{u}_i \cdot \mathbf{e}_r = \mathbf{u}_e \cdot \mathbf{e}_r = 0$,

(f4) the tangential component of the total stress is continuous at the drop surface; at $r = 1$, $\mathbf{e}_r \cdot \boldsymbol{\tau}_i \cdot \left(\mathbf{I} - \mathbf{e}_r\mathbf{e}_r\right) = \mathbf{e}_r \cdot \boldsymbol{\tau}_e \cdot \left(\mathbf{I} - \mathbf{e}_r\mathbf{e}_r\right)$, where $\boldsymbol{\tau}_{i,e}$ is the total stress tensor with contribution from hydrodynamic and electrical effects, and $\mathbf{I} - \mathbf{e}_r\mathbf{e}_r$ represents the surface projection operator.

## 2.2. Expansion in perturbation of $Re_E$

One important thing to note here is that the solution of the present electrohydrodynamic problem is not straight forward due to the non-linearity associated with the charge convection (see boundary condition e4) at the drop surface even in the limit of non-deforming spherical drop. In order to make further analytical development, we consider the electric Reynolds number $Re_E$ as the perturbation parameter. Here we seek a perturbation solution for the electric potential, velocity, pressure and stress fields of the following form

$$\begin{aligned}
\psi &= \psi^{(0)} + Re_E\psi^{(1)} + Re_E^2\psi^{(2)} + \cdots, \\
\mathbf{u} &= \mathbf{u}^{(0)} + Re_E\mathbf{u}^{(1)} + Re_E^2\mathbf{u}^{(2)} + \cdots, \\
p &= p^{(0)} + Re_E p^{(1)} + Re_E^2 p^{(2)} + \cdots, \\
\boldsymbol{\tau} &= \boldsymbol{\tau}^{(0)} + Re_E\boldsymbol{\tau}^{(1)} + Re_E^2\boldsymbol{\tau}^{(2)} + \cdots.
\end{aligned} \tag{2.3}$$

The unknown drop velocity $\mathbf{U}_d$ is also expanded in the following form

$$\mathbf{U}_d = \mathbf{U}_d^{(0)} + Re_E\mathbf{U}_d^{(1)} + Re_E^2\mathbf{U}_d^{(2)} + \cdots, \tag{2.4}$$

where $\mathbf{U}_d^{(0)}$ is the leading order drop velocity representing the drop velocity in the absence of charge convection, and $\mathbf{U}_d^{(1)}$ brings the first correction to drop velocity due to charge convection.

The leading order electric potential $\psi_{i,e}^{(0)}$ satisfies the Laplace equations of the form:

$$\nabla^2 \psi_{i,e}^{(0)} = 0. \tag{2.5}$$

The boundary conditions (e1-e4) at the leading order transforms to the following

$$\begin{aligned}
&\text{at } r \to \infty, \quad \psi_e^{(0)} \to \psi_\infty, \\
&\text{at } r = 0, \ \psi_i^{(0)} \text{ is bounded}, \\
&\text{at } r = 1, \ \psi_i^{(0)} = \psi_e^{(0)}, \\
&\text{at } r = 1, \quad \mathbf{e}_r \cdot \left( R \nabla \psi_i^{(0)} - \nabla \psi_e^{(0)} \right) = 0.
\end{aligned} \tag{2.6}$$

The leading order inner and outer flow field satisfy the Stokes equation and the condition of incompressibility of the form:

$$\nabla p_{i,e}^{(0)} = N_\mu \nabla^2 \mathbf{u}_{i,e}^{(0)}, \ \nabla \cdot \mathbf{u}_{i,e}^{(0)} = 0. \tag{2.7}$$

The boundary conditions (f1-f4) at the leading order are

$$\begin{aligned}
&\text{at } r \to \infty, \mathbf{u}_e^{(0)} = \mathbf{u}_\infty^{(0)} = \mathbf{V}_\infty - \mathbf{U}_d^{(0)}, \\
&\text{at } r = 0, \ \mathbf{u}_i^{(0)} \text{ is bounded}, \\
&\text{at } r = 1, \ \mathbf{u}_i^{(0)} = \mathbf{u}_e^{(0)}, \\
&\text{at } r = 1, \ \mathbf{u}_i^{(0)} \cdot \mathbf{e}_r = \mathbf{u}_e^{(0)} \cdot \mathbf{e}_r = 0, \\
&\text{at } r = 1, \ \mathbf{e}_r \cdot \boldsymbol{\tau}_i^{(0)} \cdot (\mathbf{I} - \mathbf{e}_r \mathbf{e}_r) = \mathbf{e}_r \cdot \boldsymbol{\tau}_e^{(0)} \cdot (\mathbf{I} - \mathbf{e}_r \mathbf{e}_r).
\end{aligned} \tag{2.8}$$

The governing equations and boundary conditions at the first order for the electric potential is given by:

$$\nabla^2 \psi_{i,e}^{(1)} = 0. \tag{2.9}$$

Quite naturally, the boundary conditions for the first order electric potential are not same as that of leading order. At first order, the effect of charge convection plays a big role in determining the electric potential distribution. The boundary conditions (e1-e4) at the first order take the following form

$$\begin{aligned}
&\text{at } r \to \infty, \quad \psi_e^{(1)} \to 0, \\
&\text{at } r = 0, \psi_i^{(1)} \text{ is bounded}, \\
&\text{at } r = 1, \ \psi_i^{(1)} = \psi_e^{(1)}, \\
&\text{at } r = 1, \quad \mathbf{e}_r \cdot \left( R \nabla \psi_i^{(1)} - \nabla \psi_e^{(1)} \right) = -\nabla_s \cdot \left( q^{(0)} \mathbf{V}_s^{(0)} \right),
\end{aligned} \tag{2.10}$$

where the surface charge distribution of the leading order is given by $q_s^{(0)} = \mathbf{e}_r \cdot \left( S \nabla \psi_i^{(0)} - \nabla \psi_e^{(0)} \right)$.

The velocity and pressure field at the first order satisfy the Stokes equation and the condition of incompressibility of the form:

$$\nabla p_{i,e}^{(1)} = N_\mu \nabla^2 \mathbf{u}_{i,e}^{(1)}, \quad \nabla \cdot \mathbf{u}_{i,e}^{(1)} = 0, \qquad (2.11)$$

which are subjected to the following boundary conditions

$$\begin{aligned}
&\text{at } r \to \infty, \ \mathbf{u}_e^{(1)} = \mathbf{u}_\infty^{(1)} = -\mathbf{U}_d^{(1)}, \\
&\text{at } r = 0, \ \mathbf{u}_i^{(1)} \text{ is bounded}, \\
&\text{at } r = 1, \ \mathbf{u}_i^{(1)} = \mathbf{u}_e^{(1)}, \\
&\text{at } r = 1, \ \mathbf{u}_i^{(1)} \cdot \mathbf{e}_r = \mathbf{u}_e^{(1)} \cdot \mathbf{e}_r = 0, \\
&\text{at } r = 1, \ \mathbf{e}_r \cdot \boldsymbol{\tau}_i^{(1)} \cdot (\mathbf{I} - \mathbf{e}_r \mathbf{e}_r) = \mathbf{e}_r \cdot \boldsymbol{\tau}_e^{(1)} \cdot (\mathbf{I} - \mathbf{e}_r \mathbf{e}_r).
\end{aligned} \qquad (2.12)$$

# 3. Analytical solution in the spherical limit

### 3.1. *First iteration - leading order electric field and flow field*

The leading order potential distribution is obtained by solving the Laplace equation (2.5) as

$$\begin{aligned}
\psi_i^{(0)} &= \sum_{n=0}^{\infty} \left( a_{n,m}^{(0)} \cos m\phi + \hat{a}_{n,m}^{(0)} \sin m\phi \right) r^n P_{n,m} (\cos\theta), \\
\psi_e^{(0)} &= \psi_\infty + \sum_{n=0}^{\infty} \left( b_{n,m}^{(0)} \cos m\phi + \hat{b}_{n,m}^{(0)} \sin m\phi \right) r^{-n-1} P_{n,m} (\cos\theta),
\end{aligned} \qquad (3.1)$$

where the unperturbed electric potential at infinity is given by $\psi_\infty = -r \left( E_x P_{1,1} \cos\phi + E_y P_{1,1} \sin\phi + E_z P_{1,0} \right)$. In the above solution, $P_{n,m} (\cos\theta)$ denotes the associated Legendre polynomial of degree $n$ and order $m$. For the sake of brevity, form now onwards we are not specifying explicitly the argument $\cos\theta$ of the associated Legendre polynomial. The unknown coefficients are evaluated by enforcing the boundary conditions (2.6). Upon this, the leading order solution for the electric potential is obtained as

$$\begin{aligned}
\psi_i^{(0)} &= -\frac{3E_z}{2+R} r P_{1,0} - \left( \frac{3E_x \cos\phi}{2+R} + \frac{3E_y \sin\phi}{2+R} \right) r P_{1,1}, \\
\psi_e^{(0)} &= -r \left( E_x P_{1,1} \cos\phi + E_y P_{1,1} \sin\phi + E_z P_{1,0} \right) + \frac{E_z (R-1)}{(2+R) r^2} P_{1,0} \\
&\quad + \left[ \frac{E_x (R-1) \cos\phi}{2+R} + \frac{E_y (R-1) \sin\phi}{2+R} \right] \frac{1}{r^2} P_{1,1}.
\end{aligned} \qquad (3.2)$$

Consequently, the leading order surface charge distribution is easily obtained from

$$q_s^{(0)} = \mathbf{e}_r \cdot \left( S \nabla \psi_i^{(0)} - \nabla \psi_e^{(0)} \right) \qquad \text{yielding} \qquad q_s^{(0)} = \frac{3(R-S)}{R+2} \left[ E_z P_{1,0} + \left( E_x \cos \phi + E_y \sin \phi \right) P_{1,1} \right].$$

Proceeding further, the velocity and pressure field inside the drop may be expressed as the Lamb solution (Lamb 1975; Hetsroni & Haber 1970; Happel & Brenner 1981)

$$\mathbf{u}_i^{(0)} = \sum_{n=1}^{\infty} \left[ \nabla \times \left( \mathbf{r} \chi_n^{(0)} \right) + \nabla \Phi_n^{(0)} + \frac{n+3}{2(n+1)(2n+3)\lambda} r^2 \nabla p_n^{(0)} - \frac{n}{(n+1)(2n+3)\lambda} \mathbf{r} p_n^{(0)} \right],$$

$$p_i^{(0)} = \sum_{n=1}^{\infty} p_n^{(0)}, \tag{3.3}$$

with $\mathbf{r}$ denoting the dimensionless position vector having magnitude $r$. The growing solid harmonics $\chi_n^{(0)}, \Phi_n^{(0)}$ and $p_n^{(0)}$ are of the following form (Hetsroni & Haber 1970):

$$p_n^{(0)} = \lambda r^n \sum_{m=0}^{n} \left( A_{n,m}^{(0)} \cos m\phi + \hat{A}_{n,m}^{(0)} \sin m\phi \right) P_{n,m},$$

$$\Phi_n^{(0)} = r^n \sum_{m=0}^{n} \left( B_{n,m}^{(0)} \cos m\phi + \hat{B}_{n,m}^{(0)} \sin m\phi \right) P_{n,m}, \tag{3.4}$$

$$\chi_n^{(0)} = r^n \sum_{m=0}^{n} \left( C_{n,m}^{(0)} \cos m\phi + \hat{C}_{n,m}^{(0)} \sin m\phi \right) P_{n,m}.$$

On similar lines, the leading order velocity and pressure field outside the drop may be expressed as the Lamb solution:

$$\mathbf{u}_e^{(0)} = \mathbf{u}_\infty^{(0)} + \mathbf{v}_e^{(0)}$$

$$= \mathbf{u}_\infty^{(0)} + \sum_{n=1}^{\infty} \left[ \nabla \times \left( \mathbf{r} \chi_{-n-1}^{(0)} \right) + \nabla \Phi_{-n-1}^{(0)} - \frac{n-2}{2n(2n-1)} r^2 \nabla p_{-n-1}^{(0)} + \frac{n+1}{n(2n-1)} \mathbf{r} p_{-n-1}^{(0)} \right], \tag{3.5}$$

$$p_e^{(0)} = \sum_{n=1}^{\infty} p_{-n-1}^{(0)},$$

where $\mathbf{v}_e$ represents the disturbance velocity filed external to the drop. The decaying solid harmonics $\chi_{-n-1}^{(0)}, \Phi_{-n-1}^{(0)}$ and $p_{-n-1}^{(0)}$ associated with $\mathbf{v}_e^{(0)}$ are given as

$$p_{-n-1}^{(0)} = r^{-n-1} \sum_{m=0}^{n} \left( A_{-n-1,m}^{(0)} \cos m\phi + \hat{A}_{-n-1,m}^{(0)} \sin m\phi \right) P_{n,m},$$

$$\Phi_{-n-1}^{(0)} = r^{-n-1} \sum_{m=0}^{n} \left( B_{-n-1,m}^{(0)} \cos m\phi + \hat{B}_{-n-1,m}^{(0)} \sin m\phi \right) P_{n,m}, \tag{3.6}$$

$$\chi_{-n-1}^{(0)} = r^{-n-1} \sum_{m=0}^{n} \left( C_{-n-1,m}^{(0)} \cos m\phi + \hat{C}_{-n-1,m}^{(0)} \sin m\phi \right) P_{n,m}.$$

Before attempting to obtain the unknown solid harmonics present in equations (3.3)-(3.6), we expresses the unperturbed velocity in drop reference frame $\mathbf{u}_\infty^{(0)}$ in terms of the specified quadratic flow field and unknown drop velocity using the relation $\mathbf{u}_\infty^{(0)} = \mathbf{V}_\infty - \mathbf{U}_d^{(0)}$. Towards this, $\mathbf{u}_\infty^{(0)}$ may be decomposed in terms of both decaying and growing the solid spherical harmonics using Lamb solution as

$$\mathbf{u}_\infty^{(0)} = \sum_{n=-\infty}^{\infty} \left[ \nabla \times \left( \mathbf{r} \chi_n^{\infty(0)} \right) + \nabla \Phi_n^{\infty(0)} + \frac{n+3}{2(n+1)(2n+3)} r^2 \nabla p_n^{\infty(0)} - \frac{n}{(n+1)(2n+3)} \mathbf{r} p_n^{\infty(0)} \right], (3.7)$$

with the solid harmonics represented in the following convenient form (Hetsroni & Haber 1970):

$$p_n^{\infty(0)} = \frac{2(2n+3)}{n} r^n \sum_{m=0}^{n} \left( \alpha_{n,m}^{(0)} \cos m\phi + \hat{\alpha}_{n,m}^{(0)} \sin m\phi \right) P_{n,m},$$

$$\Phi_n^{\infty(0)} = \frac{1}{n} r^n \sum_{m=0}^{n} \left( \beta_{n,m}^{(0)} \cos m\phi + \hat{\beta}_{n,m}^{(0)} \sin m\phi \right) P_{n,m}, \tag{3.8}$$

$$\chi_n^{\infty(0)} = \frac{1}{n(n+1)} r^n \sum_{m=0}^{n} \left( \gamma_{n,m}^{(0)} \cos m\phi + \hat{\gamma}_{n,m}^{(0)} \sin m\phi \right) P_{n,m}.$$

Similarly, the unperturbed imposed velocity $\mathbf{V}_\infty$ also satisfies the Stokes equation, thus can be represented by Lamb solution of the form

$$\mathbf{V}_\infty^{(0)} = \sum_{n=-\infty}^{\infty} \left[ \nabla \times \left( \mathbf{r} \chi_n^{\infty(0)} \right) + \nabla \Phi_n^{\infty(0)} + \frac{n+3}{2(n+1)(2n+3)} r^2 \nabla p_n^{\infty(0)} - \frac{n}{(n+1)(2n+3)} \mathbf{r} p_n^{\infty(0)} \right],$$

$$\tag{3.9}$$

where the solid spherical harmonics are of the form

$$p_n^{\infty(0)} = \frac{2(2n+3)}{n} r^n \sum_{m=0}^{n} \left( \zeta_{n,m}^{(0)} \cos m\phi + \hat{\zeta}_{n,m}^{(0)} \sin m\phi \right) P_{n,m},$$

$$\Phi_n^{\infty(0)} = \frac{1}{n} r^n \sum_{m=0}^{n} \left( \eta_{n,m}^{(0)} \cos m\phi + \hat{\eta}_{n,m}^{(0)} \sin m\phi \right) P_{n,m}, \tag{3.10}$$

$$\chi_n^{\infty(0)} = \frac{1}{n(n+1)} r^n \sum_{m=0}^{n} \left( \varpi_{n,m}^{(0)} \cos m\phi + \hat{\varpi}_{n,m}^{(0)} \sin m\phi \right) P_{n,m}.$$

The coefficients $\alpha_{n,m}^{(0)}, \hat{\alpha}_{n,m}^{(0)}, \beta_{n,m}^{(0)}, \hat{\beta}_{n,m}^{(0)}, \gamma_{n,m}^{(0)}$ and $\hat{\gamma}_{n,m}^{(0)}$ appearing in equation (3.8) can be related to the known coefficients $\zeta_{n,m}^{(0)}, \hat{\zeta}_{n,m}^{(0)}, \eta_{n,m}^{(0)}, \hat{\eta}_{n,m}^{(0)}, \varpi_{n,m}^{(0)}$ and $\hat{\varpi}_{n,m}^{(0)}$ appearing in equation (3.10) and the unknown drop velocity at leading order, $\mathbf{U}_d^{(0)} = U_{dx}^{(0)} \mathbf{e}_x + U_{dy}^{(0)} \mathbf{e}_y + U_{dz}^{(0)} \mathbf{e}_z$ in of the following form (see appendix A for details):

$$\beta_{1,0}^{(0)} = \eta_{1,0} - U_{dz}^{(0)},$$
$$\beta_{1,1}^{(0)} = \eta_{1,1} - U_{dx}^{(0)}, \qquad (3.11)$$
$$\hat{\beta}_{1,1}^{(0)} = \hat{\eta}_{1,1} - U_{dy}^{(0)},$$

and for all other values of $n$ and $m$ we obtain

$$\left\{ \alpha_{n,m}^{(0)}, \alpha_{-n-1,m}^{(0)}, \hat{\alpha}_{n,m}^{(0)}, \hat{\alpha}_{-n-1,m}^{(0)} \right\} = \left\{ \zeta_{n,m}, \zeta_{-n-1,m}, \hat{\zeta}_{n,m}, \hat{\zeta}_{-n-1,m} \right\},$$

$$\left\{ \beta_{n,m}^{(0)}, \beta_{-n-1,m}^{(0)}, \hat{\beta}_{n,m}^{(0)}, \hat{\beta}_{-n-1,m}^{(0)} \right\} = \left\{ \eta_{n,m}, \eta_{-n-1,m}, \hat{\eta}_{n,m}, \hat{\eta}_{-n-1,m} \right\}, \qquad (3.12)$$

$$\left\{ \gamma_{n,m}^{(0)}, \gamma_{-n-1,m}^{(0)}, \hat{\gamma}_{n,m}^{(0)}, \hat{\gamma}_{-n-1,m}^{(0)} \right\} = \left\{ \varpi_{n,m}, \varpi_{-n-1,m}, \hat{\varpi}_{n,m}, \hat{\varpi}_{-n-1,m} \right\}.$$

Now, the coefficients $A_{n,m}^{(0)}, B_{n,m}^{(0)}, C_{n,m}^{(0)}, A_{-n-1,m}^{(0)}, B_{-n-1,m}^{(0)}$ and $C_{-n-1,m}^{(0)}$ are obtained by enforcing the boundary conditions (2.8). Following the convection (Happel & Brenner 1981; Hetsroni & Haber 1970; Hetsroni et al. 1970), we may cast the boundary conditions as

$$\left[ \mathbf{u}_{i,r}^{(0)} \cdot \mathbf{e}_r \right]_s = 0,$$

$$\left[ \mathbf{v}_e^{(0)} \cdot \mathbf{e}_r \right]_s + \left[ \mathbf{u}_\infty^{(0)} \cdot \mathbf{e}_r \right]_s = 0,$$

$$\left[ r \frac{\partial}{\partial r} \left( \mathbf{u}_i^{(0)} \cdot \mathbf{e}_r \right) \right]_s = \left[ r \frac{\partial}{\partial r} \left( \mathbf{u}_\infty^{(0)} \cdot \mathbf{e}_r \right) \right] + \left[ r \frac{\partial}{\partial r} \left( \mathbf{v}_e^{(0)} \cdot \mathbf{e}_r \right) \right],$$

$$\left[ \mathbf{r} \cdot \nabla \times \mathbf{u}_i^{(0)} \right]_s = \left[ \mathbf{r} \cdot \nabla \times \mathbf{u}_\infty^{(0)} \right]_s + \left[ \mathbf{r} \cdot \nabla \times \mathbf{v}_e^{(0)} \right]_s,$$

$$\left[ \mathbf{r} \cdot \nabla \times \left\{ \mathbf{r} \times \left( \left( \boldsymbol{\tau}_i^{H(0)} + M \boldsymbol{\tau}_i^{E(0)} \right) \cdot \mathbf{e}_r \right) \right\} \right]_s = \left[ \mathbf{r} \cdot \nabla \times \left\{ \mathbf{r} \times \left( \boldsymbol{\tau}_\infty^{H(0)} \cdot \mathbf{e}_r \right) \right\} \right]_s + \left[ \mathbf{r} \cdot \nabla \times \left\{ \mathbf{r} \times \left( \left( \boldsymbol{\tau}_e^{H(0)} + M \boldsymbol{\tau}_e^{E(0)} \right) \cdot \mathbf{e}_r \right) \right\} \right]_s,$$

$$\left[ \mathbf{r} \cdot \nabla \times \left\{ \left( \boldsymbol{\tau}_i^{H(0)} + M \boldsymbol{\tau}_i^{E(0)} \right) \cdot \mathbf{e}_r \right\} \right]_s = \left[ \mathbf{r} \cdot \nabla \times \left( \boldsymbol{\tau}_\infty^{H(0)} \cdot \mathbf{e}_r \right) \right]_s + \left[ \mathbf{r} \cdot \nabla \times \left\{ \left( \boldsymbol{\tau}_e^{H(0)} + M \boldsymbol{\tau}_e^{E(0)} \right) \cdot \mathbf{e}_r \right\} \right]_s,$$

$$(3.13)$$

with the $\left[ \vartheta \right]_r$ denoting the evaluation of the generic variable $\vartheta$ at the drop interface i.e. $r = 1$. The expressions of the hydrodynamic stresses and the electrical Maxwell stresses at the leading order appearing above are given as

$$\boldsymbol{\tau}_i^{H(0)} = \lambda\left[-p_i^{(0)}\mathbf{I} + \nabla\mathbf{u}_i^{(0)} + \left(\nabla\mathbf{u}_i^{(0)}\right)^T\right],$$

$$\boldsymbol{\tau}_e^{H(0)} = \left[-p_e^{(0)}\mathbf{I} + \nabla\mathbf{v}_e^{(0)} + \left(\nabla\mathbf{v}_e^{(0)}\right)^T\right],$$

$$\boldsymbol{\tau}_\infty^{H(0)} = \left[-p_\infty^{(0)}\mathbf{I} + \nabla\mathbf{u}_\infty^{(0)} + \left(\nabla\mathbf{u}_\infty^{(0)}\right)^T\right],$$

$$\boldsymbol{\tau}_i^{E(0)} = S\left[\mathbf{E}_i^{(0)}\left(\mathbf{E}_i^{(0)}\right)^T - \frac{1}{2}\left|\mathbf{E}_i^{(0)}\right|^2\mathbf{I}\right],$$

$$\boldsymbol{\tau}_e^{E(0)} = \left[\mathbf{E}_e^{(0)}\left(\mathbf{E}_e^{(0)}\right)^T - \frac{1}{2}\left|\mathbf{E}_e^{(0)}\right|^2\mathbf{I}\right],$$

where superscript $T$ is used to denote the transpose. Referring to appendix B, the form of the coefficients are obtained as

$$A_{n,m}^{(0)} = \frac{(2n+3)}{n(2n+1)(\lambda+1)}\left[\left(4n^2+8n+3\right)\alpha_{n,m}^{(0)} + \left(4n^2-1\right)\beta_{n,m}^{(0)} + M\left(g_{n,m}^{i(0)} - g_{n,m}^{e(0)}\right)\right],$$

$$A_{-n-1,m}^{(0)} = -\frac{1}{(n+1)(2n+1)(\lambda+1)}\left[\lambda\left(8n^3+12n^2-2n-3\right)\alpha_{n,m}^{(0)} + (1+\lambda)\left(8n^2-2\right)\alpha_{-n-1,m}^{(0)}\right.$$
$$\left. + \left\{\lambda\left(8n^2-2\right) + \left(8n^3+4n^2-2n-1\right)\right\}\beta_{n,m}^{(0)} + M\left(1-2n\right)\left(g_{n,m}^{i(0)} - g_{n,m}^{e(0)}\right)\right],$$

$$B_{n,m}^{(0)} = -\frac{A_{n,m}^{(0)}}{2(2n+3)},$$

$$B_{-n-1,m}^{(0)} = -\frac{1}{2\lambda\left(2n^2+3n+1\right)}\left[\left\{\lambda\left(4n^2+4n+1\right) - (4n+2)\right\}\alpha_{n,m}^{(0)} + \lambda\left(4n^2-1\right)\beta_{n,m}^{(0)}\right.$$
$$\left. - (1+\lambda)(4n+2)\beta_{-n-1,m}^{(0)} - M\left(g_{n,m}^{i(0)} - g_{n,m}^{e(0)}\right)\right],$$

$$C_{n,m}^{(0)} = \frac{(2n+1)\gamma_{n,m}^{(0)} + M\left(h_{n,m}^{e(0)} - h_{n,m}^{i(0)}\right)}{n(n+1)(n+2+\lambda n - \lambda)},$$

$$C_{-n-1,m}^{(0)} = \frac{\left[(n-1)(1-\lambda)\gamma_{n,m}^{(0)} + \left\{(2+n)+\lambda(n-1)\right\}\gamma_{-n-1,m}^{(0)} + M\left(h_{n,m}^{e(0)} - h_{n,m}^{i(0)}\right)\right]}{n\left\{(n^2+3n+2)+\lambda(n^2-1)\right\}}.$$

$$(3.14)$$

In the above expressions, the coefficients $g_{n,m}^{i,e(0)}$, $\hat{g}_{n,m}^{i,e(0)}$, $h_{n,m}^{i,e(0)}$ and $\hat{h}_{n,m}^{i,e(0)}$ are obtained from the suitable surface harmonic representation of the electrical stresses as depicted in Appendix B, equations (B7)-(B8). One thing to note here is that the expressions of $\hat{A}_{n,m}^{(0)}, \hat{B}_{n,m}^{(0)}, \hat{C}_{n,m}^{(0)}, \hat{A}_{-n-1,m}^{(0)}, \hat{B}_{-n-1,m}^{(0)}$ and $\hat{C}_{-n-1,m}^{(0)}$ are obtained by replacing the terms $\alpha_{n,m}^{(0)}, \alpha_{-n-1,m}^{(0)}, \beta_{n,m}^{(0)}, \beta_{-n-1,m}^{(0)}, g_{n,m}^{i(0)}$ and $h_{n,m}^{i(0)}$ by $\hat{\alpha}_{n,m}^{(0)}, \hat{\alpha}_{-n-1,m}^{(0)}, \hat{\beta}_{n,m}^{(0)}, \hat{\beta}_{-n-1,m}^{(0)}, \hat{g}_{n,m}^{i(0)}$ and $\hat{h}_{n,m}^{i(0)}$, respectively. The complete expressions of the velocity field and the pressure field inside and outside the drop at the leading order are given in Appendix C.

## 3.2. *Leading order drop velocity: force free condition*

The drop velocity is determined from the force free condition (Hetsroni & Haber 1970). The total force acting on the drop is given by

$$\mathbf{F} = \mathbf{F}^H + M\mathbf{F}^E = \int_A \left( \boldsymbol{\tau}_e^H \cdot \mathbf{e}_r \right) dA + M \int_A \left( \boldsymbol{\tau}_e^E \cdot \mathbf{e}_r \right) dA$$
$$= -4\pi \nabla \left( r^3 p_{-2} \right) + M \int_A \left( \boldsymbol{\tau}_e^E \cdot \mathbf{e}_r \right) dA, \tag{3.15}$$

where the integration is performed on the spherical drop surface. Now, using the perturbation expansion (equation (2.3)), at the leading order of approximation we obtain

$$\mathbf{F}^{(0)} = \mathbf{F}^{H(0)} + M\mathbf{F}^{E(0)} = -4\pi \nabla \left( r^3 p_{-2}^{(0)} \right) + M \int_A \left( \boldsymbol{\tau}_e^{E(0)} \cdot \mathbf{e}_r \right) dA. \tag{3.16}$$

At leading order of approximation the electrical force on the drop is identically zero, therefore the force free condition at leading order is given by

$$\nabla \left( r^3 p_{-2}^{(0)} \right) = \mathbf{0}, \tag{3.17}$$

where $p_{-2}^{(0)} = r^{-2} \left[ A_{-2,0}^{(0)} P_{1,0} + \left( A_{-2,1}^{(0)} \cos\phi + \hat{A}_{-2,1}^{(0)} \sin\phi \right) P_{1,1} \right]$. After substitution of the expressions of $A_{-2}^0, A_{-2}^1$ and $\hat{A}_{-2}^1$ from equation (3.14), the above vector equation yields three scalar equations of the form

$$15\lambda \alpha_{1,0}^{i(0)} + 9\lambda \beta_{1,0}^{(0)} + 6\beta_{1,0}^{(0)} - Mg_{1,0}^{i(0)} + Mg_{1,0}^{e(0)} = 0,$$
$$15\lambda \alpha_{1,1}^{i(0)} + 9\lambda \beta_{1,1}^{(0)} + 6\beta_{1,1}^{(0)} - Mg_{1,1}^{i(0)} + Mg_{1,1}^{e(0)} = 0, \tag{3.18}$$
$$15\lambda \hat{\alpha}_{1,1}^{i(0)} + 9\lambda \hat{\beta}_{1,1}^{(0)} + 6\hat{\beta}_{1,1}^{(0)} - M\hat{g}_{1,1}^{i(0)} + M\hat{g}_{1,1}^{e(0)} = 0.$$

We note that the hydrodynamic force due to the unperturbed pressure field is zero i.e. $\nabla \left( r^3 p_{-2}^\infty \right) = 0$, which yields $\alpha_{-2}^0 = \alpha_{-2}^1 = \hat{\alpha}_{-2}^1 = 0$. Towards determining the drop velocity we substitute $\alpha, \beta$ and $\gamma$ in terms of known coefficients $\zeta, \eta, \varpi$ and unknown drop velocity using the relations (3.11) and (3.12). Finally, the drop velocity is obtained as

$$U_{dx} = \frac{15\lambda \zeta_{1,1} + (9\lambda + 6)\eta_{1,1} + M \left( g_{1,1}^{e(0)} - g_{1,1}^{i(0)} \right)}{3(3\lambda + 2)},$$

$$U_{dy} = \frac{15\lambda \hat{\zeta}_{1,1} + (9\lambda + 6)\hat{\eta}_{1,1} + M \left( \hat{g}_{1,1}^{e(0)} - \hat{g}_{1,1}^{i(0)} \right)}{3(3\lambda + 2)}, \tag{3.19}$$

$$U_{dz} = \frac{15\lambda \zeta_{1,0} + (9\lambda + 6)\eta_{1,0} + M \left( g_{1,0}^{e(0)} - g_{1,0}^{i(0)} \right)}{3(3\lambda + 2)}.$$

Now, substituting $\zeta_{1,0} = (k_2 + k_3)/5$, $\eta_{1,0} = k_0$, $\zeta_{1,1} = \hat{\zeta}_{1,1} = 0$, $\eta_{1,1} = \hat{\eta}_{1,1} = 0$, $g_{1,0}^{i(0)} = g_{1,0}^{e(0)} = 0$, $g_{1,1}^{i(0)} = g_{1,1}^{e(0)} = 0$ and $\hat{g}_{1,1}^{i(0)} = \hat{g}_{1,1}^{e(0)} = 0$ (see Appendix A and B for detail) in the above expressions and obtain the drop velocity at the leading order as

$$
\begin{aligned}
U_{dx}^{(0)} &= U_{dy}^{(0)} = 0, \\
U_{dz}^{(0)} &= \frac{(3k_0 + 2k_2)\lambda + 2k_0}{3\lambda + 2}.
\end{aligned}
\tag{3.20}
$$

### 3.3. Second iteration - effect of surface convection on the electric field

After obtaining the electric and flow field at the leading order, we now obtain the solution to equations (2.9) and (2.11) using the boundary conditions (2.10) and (2.12), respectively. The effect of charge convection is apparent from the given boundary condition $\mathbf{e}_r \cdot \left( R \nabla \psi_i^{(1)} - \nabla \psi_e^{(1)} \right) = -\nabla_s \cdot \left( q^{(0)} \mathbf{V}_s^{(0)} \right)$, the leading order charge distribution and surface velocity are responsible for the alteration in the first order charge conservation equation at the drop surface.

The solution of the Laplace equation for the first order electric potential are similarly obtained as:

$$
\begin{aligned}
\psi_i^{(1)} &= \sum_{n=0}^{\infty} \left( a_{n,m}^{(1)} \cos m\phi + \hat{a}_{n,m}^{(1)} \sin m\phi \right) r^n P_{n,m}(\cos\theta), \\
\psi_e^{(1)} &= \sum_{n=0}^{\infty} \left( b_{n,m}^{(1)} \cos m\phi + \hat{b}_{n,m}^{(1)} \sin m\phi \right) r^{-n-1} P_{n,m}(\cos\theta).
\end{aligned}
\tag{3.21}
$$

To obtain the non-zero spherical harmonics present in the first order electric potential, we first evaluate the term $\nabla_s \cdot \left( q^{(0)} \mathbf{V}_s^{(0)} \right)$ in the following form (Kim & Karrila 1991)

$$
\nabla_s \cdot \left( q^{(0)} \mathbf{V}_s^{(0)} \right) = 2 \left( q^{(0)} \mathbf{V}_s^{(0)} \cdot \mathbf{e}_r \right) + \frac{1}{\sin\theta} \frac{\partial}{\partial\theta} \left( q^{(0)} \mathbf{V}_s^{(0)} \cdot \mathbf{e}_\theta \right) + \frac{1}{\sin\theta} \frac{\partial}{\partial\phi} \left( q^{(0)} \mathbf{V}_s^{(0)} \cdot \mathbf{e}_\phi \right).
\tag{3.22}
$$

Now, we substitute the expressions of the velocity field (given in appendix C) and the surface charge distribution $\left( q_s^{(0)} \right)$ in equation (3.22). We represent the right hand side of equation (3.22) in terms of various surface harmonics by using the orthogonality of the spherical surface harmonics as:

$$
\begin{aligned}
-2 \left( q^{(0)} \mathbf{V}_s^{(0)} \cdot \mathbf{e}_r \right) &- \frac{1}{\sin\theta} \frac{\partial}{\partial\theta} \left( q^{(0)} \mathbf{V}_s^{(0)} \cdot \mathbf{e}_\theta \right) - \frac{1}{\sin\theta} \frac{\partial}{\partial\phi} \left( q^{(0)} \mathbf{V}_s^{(0)} \cdot \mathbf{e}_\phi \right) \\
&= \sum_{n=0}^{\infty} \sum_{m=0}^{n} \left( Z_{n,m} \cos m\phi + \hat{Z}_{n,m} \sin m\phi \right) P_{n,m}.
\end{aligned}
\tag{3.23}
$$

| Non-zero coefficients of $\psi_i^{(1)}$ | Non-zero coefficients of $\psi_e^{(1)}$ |
|---|---|
| $a_{1,0}^{(1)}, a_{1,1}^{(1)}, \hat{a}_{1,1}^{(1)}$ | $b_{-2,0}^{(1)}, b_{-2,1}^{(1)}, \hat{b}_{-2,1}^{(1)}$ |
| $a_{2,0}^{(1)}, a_{2,1}^{(1)}, \hat{a}_{2,1}^{(1)}, a_{2,2}^{(1)}$ | $b_{-3,0}^{(1)}, b_{-3,1}^{(1)}, \hat{b}_{-3,1}^{(1)}, b_{-3,2}^{(1)}$ |
| $a_{3,0}^{(1)}, a_{3,1}^{(1)}, \hat{a}_{3,1}^{(1)}, a_{3,2}^{(1)}, \hat{a}_{3,2}^{(1)}, a_{3,3}^{(1)}, \hat{a}_{3,3}^{(1)}$ | $b_{-4,0}^{(1)}, b_{-4,1}^{(1)}, \hat{b}_{-4,1}^{(1)}, b_{-4,2}^{(1)}, \hat{b}_{-4,2}^{(1)}, \hat{b}_{-4,3}^{(1)}, \hat{b}_{-4,3}^{(1)}$ |
| $a_{4,0}^{(1)}, a_{4,1}^{(1)}, \hat{a}_{4,1}^{(1)}, a_{4,2}^{(1)}, a_{4,3}^{(1)}, \hat{a}_{4,3}^{(1)}$ | $b_{-5,0}^{(1)}, b_{-5,1}^{(1)}, \hat{b}_{-5,1}^{(1)}, b_{-5,2}^{(1)}, \hat{b}_{-5,3}^{(1)}, \hat{b}_{-5,3}^{(1)}$ |

TABLE 1. Non-zero coefficients present in the first order electric potential.

The non-zero coefficients are: $Z_{1,0}, Z_{1,1}, Z_{2,0}, Z_{2,1}, Z_{2,2}, Z_{3,0}, Z_{3,1}, Z_{3,2}, Z_{3,3}, Z_{4,0}, Z_{4,1}, Z_{4,2}, Z_{4,3}, \hat{Z}_{1,1}, \hat{Z}_{2,1}, \hat{Z}_{3,1}, \hat{Z}_{3,2}, \hat{Z}_{3,3}, \hat{Z}_{4,1}$ and $\hat{Z}_{4,2}$ (see Appendix D for complete expression of Z's). These non-zero surface harmonics suggest that the electric potential at the first order should contain the non-zero coefficients $a_{n,m}^{(1)}, \hat{a}_{n,m}^{(1)}, b_{-n-1,m}^{(1)}$ and $\hat{b}_{-n-1,m}^{(1)}$ with $n$ and $m$ given in table 1. The detailed solution and the expressions of the coefficients $a_{n,m}^{(1)}, \hat{a}_{n,m}^{(1)}, b_{-n-1,m}^{(1)}$ and $\hat{b}_{-n-1,m}^{(1)}$ are given in Appendix E.

### 3.4. $O\left(Re_E\right)$ drop velocity

The first order velocity field and pressure field satisfy the Stokes equation (2.11) in terms of the solid spherical harmonics $p_{n,m}^{(1)}, \Phi_{n,m}^{(1)}, \chi_{n,m}^{(1)}, p_{-n-1,m}^{(1)}, \Phi_{-n-1,m}^{(1)}$ and $\chi_{-n-1,m}^{(1)}$ as obtained through the Lamb solution. The details of the workings can be found in Appendix F. The drop velocity can be obtained using the force free condition at the first order as:

$$\mathbf{F}^{(1)} = -4\pi\nabla\left(r^3 p_{-2}^{(1)}\right) = \mathbf{0}. \tag{3.24}$$

Now, substituting $p_{-2}^{(1)} = r^{-2}\left[A_{-2,0}^{(1)}P_{1,0} + \left(A_{-2,1}^{(1)}\cos\phi + \hat{A}_{-2,1}^{(1)}\sin\phi\right)P_{1,1}\right]$ in the above equation, we obtain the following scalar equations

$$A_{-2,0}^{(1)} = 0, A_{-2,1}^{(1)} = 0, \hat{A}_{-2,1}^{(1)} = 0. \tag{3.25}$$

This yields the first order drop velocity:

$$U_{dx}^{(1)} = \frac{M\left(g_{1,1}^{e(1)} - g_{1,1}^{i(1)}\right)}{3\left(2+3\lambda\right)}, \ U_{dy}^{(1)} = \frac{M\left(\hat{g}_{1,1}^{e(1)} - \hat{g}_{1,1}^{i(1)}\right)}{3\left(2+3\lambda\right)}, \ U_{dz}^{(1)} = \frac{M\left(g_{1,0}^{e(1)} - g_{1,0}^{i(1)}\right)}{3\left(2+3\lambda\right)}. \tag{3.26}$$

Substituting the expressions of the coefficients $g_{n,m}^{e,i(1)}$ and $\hat{g}_{n,m}^{e,i(1)}$ (see Appendix F, equation (F9) for detail) in equation (3.26), we obtain

$$U_{dx}^{(1)} = \frac{6}{35} \frac{ME_x E_z (R-S)(3R-S+3)\left[\left(2\lambda^2 + 63\lambda + 45\right)k_2 + \left(17\lambda^2 + 28\lambda + 15\right)k_3\right]}{\left(2+3\lambda\right)^2 \left(\lambda+1\right)\left(4+\lambda\right)\left(R+2\right)^2 \left(3+2R\right)},$$

$$U_{dy}^{(1)} = \frac{6}{35} \frac{ME_y E_z (R-S)(3R-S+3)\left[\left(17\lambda^2 + 28\lambda + 15\right)k_2 + \left(2\lambda^2 + 63\lambda + 45\right)k_3\right]}{\left(2+3\lambda\right)^2 \left(\lambda+1\right)\left(4+\lambda\right)\left(R+2\right)^2 \left(3+2R\right)},$$

$$U_{dz}^{(1)} = \frac{6}{35} \frac{M(R-S)(3R-S+3)}{\left(2+3\lambda\right)^2 \left(\lambda+1\right)\left(4+\lambda\right)\left(R+2\right)^2 \left(3+2R\right)}\Big[\big\{\left(36\lambda^2 + 119\lambda + 75\right)k_2$$
$$+\left(-9\lambda^2 + 14\lambda + 25\right)k_3\big\}E_x^2 + \big\{\left(-9\lambda^2 + 14\lambda + 25\right)k_2 + \left(36\lambda^2 + 119\lambda + 75\right)k_3\big\}E_y^2$$
$$+\left(8\lambda^2 + 42\lambda + 40\right)\left(k_2 + k_3\right)E_z^2\Big].$$

$$(3.27)$$

Following this iterative method, the higher order corrections to the drop velocity can be obtained.

## 4. Summary and discussion

In this section, we consider a special case of an unbounded cylindrical Poiseuille flow to illustrate the important implications of the electric field and more specifically the effect of surface charge convection. The imposed velocity at infinity is of the form (Hetsroni & Haber 1970; Pak et al. 2014)

$$\mathbf{V}_\infty = \left[1 - \left(\frac{r}{R_0}\right)^2 \sin^2\theta - \left(\frac{b}{R_0}\right)^2 - \frac{2rb}{R_0^2}\sin\theta\cos\phi\right]\mathbf{e}_z, \qquad (4.1)$$

where $b$ is the distance of the drop centre from the centreline of the imposed velocity and $R_0$ is the distance from the position of maximum imposed velocity to the position of zero imposed velocity. This velocity profile can be obtained by considering $k_0 = 1 - \left(\frac{b}{R_0}\right)^2$, $k_1 = -\frac{2b}{R_0^2}$ and $k_2 = k_3 = -\frac{1}{R_0^2}$ in the expression for the general quadratic flow field. Now, substituting this form of the velocity field in expressions (3.20) and (3.27) we obtain the drop migration velocity as

$$\mathbf{U}_d = \mathbf{U}_d^{(0)} + Re_E \mathbf{U}_d^{(1)} + O\left(Re_E^2\right), \qquad (4.2)$$

where the leading order migration velocity is obtained as

$$U_{dx}^{(0)} = 0,$$

$$U_{dy}^{(0)} = 0,$$

$$U_{dz}^{(0)} = \left[ 1 - \left( \frac{b}{R_0} \right)^2 - \frac{2\lambda}{R_0^2 \left( 2 + 3\lambda \right)} \right].$$

(4.3)

This leading order drop velocity is the same as the velocity of a spherical drop in an unbounded cylindrical Poiseuille flow which has been previously obtained by Haber & Hetsroni (1970). At this order of approximation the applied electric field only affects the flow field in and around the drop, but the drag force on the drop remains unaltered which leads to no change in the drop velocity. The first correction to the drop velocity is obtained using equation (3.27) as

$$U_{dx}^{(1)} = - \left[ \frac{6}{35} \frac{M E_x E_z \left( R - S \right) \left( 3R - S + 3 \right) \left( 19\lambda^2 + 91\lambda + 60 \right)}{R_0^2 \left( R + 2 \right)^2 \left( 3 + 2R \right) \left( 2 + 3\lambda \right)^2 \left( \lambda + 1 \right) \left( 4 + \lambda \right)} \right],$$

$$U_{dy}^{(1)} = - \left[ \frac{6}{35} \frac{M E_y E_z \left( R - S \right) \left( 3R - S + 3 \right) \left( 19\lambda^2 + 91\lambda + 60 \right)}{R_0^2 \left( R + 2 \right)^2 \left( 3 + 2R \right) \left( 2 + 3\lambda \right)^2 \left( \lambda + 1 \right) \left( 4 + \lambda \right)} \right],$$

$$U_{dz}^{(1)} = - \left[ \frac{6}{35} \frac{M \left( R - S \right) \left( 3R - S + 3 \right) \left\{ \left( 27\lambda^2 + 133\lambda + 100 \right) \left( E_x^2 + E_y^2 \right) + \left( 16\lambda^2 + 84\lambda + 80 \right) E_z^2 \right\}}{R_0^2 \left( R + 2 \right)^2 \left( 3 + 2R \right) \left( 2 + 3\lambda \right)^2 \left( \lambda + 1 \right) \left( 4 + \lambda \right)} \right].$$

(4.4)

The first correction of the drop velocity due to charge convection clearly depicts the cross-stream migration $\left( \text{non-zero } U_{dx}^{(1)} \text{ and/or } U_{dy}^{(1)} \right)$ of the drop along with the alteration in axial velocity of the drop $\left( \text{non-zero } U_{dz}^{(1)} \right)$. The important thing that can be understood from the expressions given in equation (4.4) is that the cross-stream migration of the drop in $x$ (or $y$) direction can take place only in the presence of *both* non-zero $E_x$ (or $E_y$) and $E_z$ components of the applied electric field. The application of a uniform electric field having only a single component is insufficient to cause cross-stream migration of a spherical drop in the presence of background cylindrical Poiseuille flow. To demonstrate this fact in more detail, we consider three different cases as   Case I: $\mathbf{E}_\infty = \left( 0, 0, 1 \right)$ (longitudinal electric field), Case II: $\mathbf{E}_\infty = \left( 1, 0, 0 \right)$ (transverse electric field) and Case III: $\mathbf{E}_\infty = \left( 1/\sqrt{2}, 0, 1/\sqrt{2} \right)$ (both longitudinal and transverse electric field).

### 4.1. *Case I: Longitudinal electric field*

In figure 2, we show the charge distribution on the drop surface in the presence of a longitudinal electric field $\left( \mathbf{E}_\infty = \mathbf{e}_z \right)$ while the background flow is considered as cylindrical Poiseuille flow. The parameters used to calculate the charge distribution are given in the figure caption. The leading order charge distribution $q^{(0)}$ shown in figure 2(a) is anti-symmetric with respect to the equatorial plane. Hence, the flow field generated due to this anti-symmetric charge distribution in the presence of $\mathbf{E}_\infty$ is also anti-symmetric which leads to zero drag force on the drop due to applied electric field. However, the surface charge convection in the presence of a Poiseuille flow markedly alters the charge distribution on the drop surface as shown in figure 2(b-c). Figure 2(b) depicts the charge distribution $q^{(0)} + Re_E q^{(1)}$ when the drop is at the centreline ($b$=0) of the background Poiseuille flow. In this case, the asymmetry in the charge distribution across the equatorial plane does not lead to any electrical force on the drop but the flow field generated by this asymmetric charge distribution alters the hydrodynamic drag force on the drop and subsequent alteration in drop velocity in $z$ direction only. However, when the drop is at an offset distance $b$=2 from the centreline, the charge distribution across both the equatorial plane and the meridional plane become asymmetric. The asymmetry in the $z$ direction leads to change in drop velocity in the axial direction as

$$\mathbf{U}_d = \left[ 1 - \left( \frac{b}{R_0} \right)^2 - \frac{2\lambda}{R_0^2 (2+3\lambda)} \right.$$
$$\left. - Re_E \left( \frac{6}{35} \frac{M(R-S)(3R-S+3)(16\lambda^2+84\lambda+80)}{R_0^2 (R+2)^2 (3+2R)(2+3\lambda)^2 (\lambda+1)(4+\lambda)} \right) \right] \mathbf{e}_z. \quad (4.5)$$

Notably, the asymmetry in the charge distribution in the $x$ direction leads to a non-zero electrical torque. The general form of the electric torque acting on the drop is obtained by taking the moment of the external electric traction vector $\boldsymbol{\tau}_e^E \cdot \mathbf{e}_r$ with respect to drop centre and integrating over the spherical drop interface as

$$\mathbf{T}^E = \int_A \mathbf{e}_r \times \left( \boldsymbol{\tau}_e^E \cdot \mathbf{e}_r \right) dA$$
$$= T_x^E \mathbf{e}_x + T_y^E \mathbf{e}_y + T_z^E \mathbf{e}_z, \quad (4.6)$$

where the components of the torque vector are obtained as

$$T_x^E = \frac{12\pi}{5} \frac{Re_E E_x E_y (R-S)(8+5\lambda)b}{R_0^2 (R+2)^2 (\lambda+1)},$$
$$T_y^E = -\frac{12\pi}{5} \frac{Re_E (R-S)\left[ (8+5\lambda)E_x^2 + (2+5\lambda)E_z^2 \right]b}{R_0^2 (R+2)^2 (\lambda+1)}, \quad (4.7)$$
$$T_z^E = \frac{12\pi}{5} \frac{Re_E E_z E_y (R-S)(8+5\lambda)b}{R_0^2 (R+2)^2 (\lambda+1)}.$$

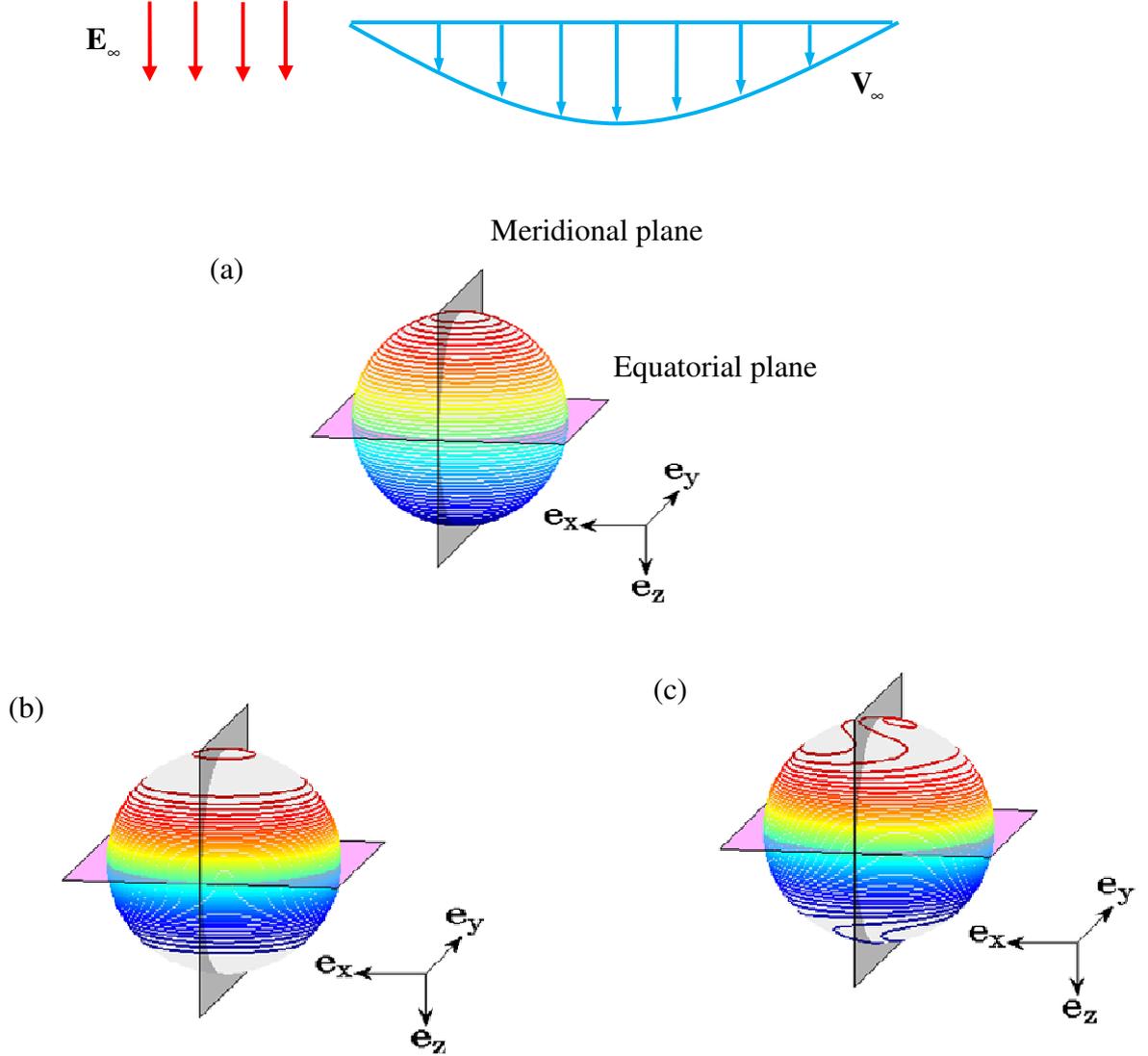

FIGURE 2. (Colour online) Contours depicting the variation of (a) $q^{(0)}$, (b) $q^{(0)} + Re_E q^{(1)}$ with $b$=0 and (c) $q^{(0)} + Re_E q^{(1)}$ with $b$=2 for a longitudinal electric field (case I). The parameters employed are $R = 0.75$, $S = 2$, $M = 2$, $\lambda = 0.25$, $Re_E = 0.2$ and $R_0 = 5$. The direction of the imposed electric field and velocity field is shown at the top.

For the Case I, substituting $E_x = E_y = 0$ and $E_z = 1$ in the above expressions given the electric torque as

$$\mathbf{T}^E = -\left[ \frac{12\pi}{5} \frac{Re_E (R - S)(2 + 5\lambda) b}{R_0^2 (R + 2)^2 (\lambda + 1)} \right] \mathbf{e}_y. \qquad (4.8)$$

An important thing to note here is that the direction of the electric torque is determined by the $R/S$ ratio and $b$. This electric torque is exactly balanced by the torque due to the hydrodynamic stresses which causes the droplet to be torque-free

$$\mathbf{T}^H + M\mathbf{T}^E = \mathbf{0}, \tag{4.9}$$

where the hydrodynamic torque acting on the drop is obtained from (Happel & Brenner 1981) $\mathbf{T}^H = -8\pi\nabla\left(r^3\chi_{-2}\right)$.

### 4.2. *Case II: Transverse electric field*

In the presence of a transverse electric field $\mathbf{E}_\infty = \mathbf{e}_x$, the charge distribution on the drop surface is shown in figure 3. Contrary to case I, here we do not obtain change in the drop velocity in the direction of applied external electric field. The effect of surface charge convection leads to asymmetric charge distribution across the equatorial plane which leads to alteration in drop velocity in the axial direction as

$$\mathbf{U}_d = \left[1-\left(\frac{b}{R_0}\right)^2 - \frac{2\lambda}{R_0^2\left(2+3\lambda\right)}\right.$$
$$\left. -Re_E\left(\frac{6}{35}\frac{M\left(R-S\right)\left(3R-S+3\right)\left(27\lambda^2+133\lambda+100\right)}{R_0^2\left(R+2\right)^2\left(3+2R\right)\left(2+3\lambda\right)^2\left(\lambda+1\right)\left(4+\lambda\right)}\right)\right]\mathbf{e}_z. \tag{4.10}$$

The charge distribution across the meridional plane remains anti-symmetric for $b=0$ (see figure 3(b)). However, for the case of $b=2$, the asymmetry in the charge distribution across the meridional plane (see figure 3(c)) leads to the generation of the electric torque in the $y$ direction as

$$\mathbf{T}^E = -\left[\frac{12\pi}{5}\frac{Re_E\left(R-S\right)\left(8+5\lambda\right)b}{R_0^2\left(R+2\right)^2\left(\lambda+1\right)}\right]\mathbf{e}_y \tag{4.11}$$

which is balanced by the hydrodynamic torque so as to satisfy the torque free condition.

### 4.3. *Case III: Combined longitudinal and transverse electric field*

The charge distribution due to the application of uniform electric field having both longitudinal and transverse components is shown in figure 4. In this situation, the charge distribution becomes asymmetric across both meridional plane and equatorial plane even for the case of $b=0$ (see figure 4(b)). This leads to the cross-stream migration of the drop in the $x$ direction along with alteration in axial velocity of the form

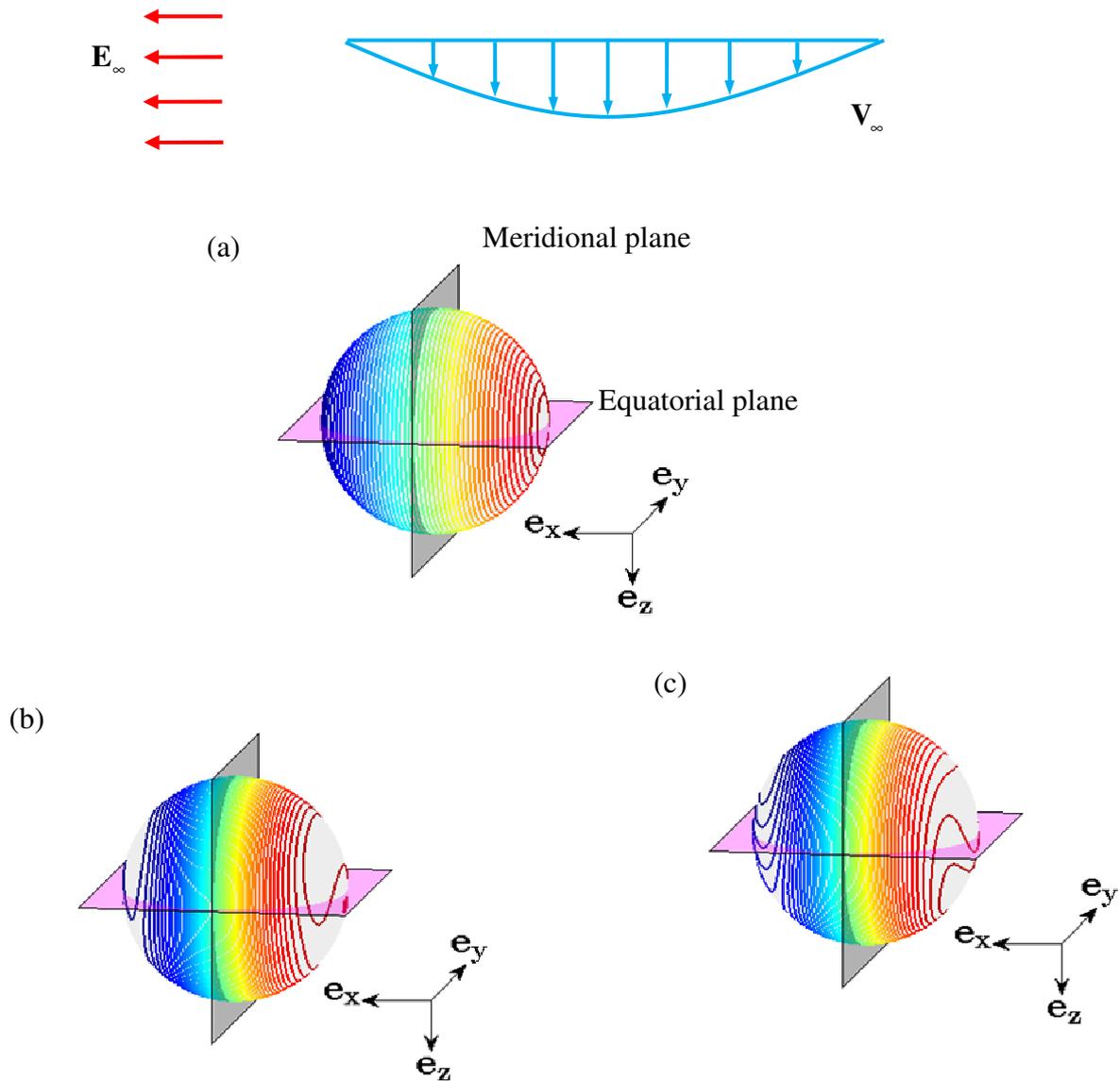

FIGURE 3. (Colour online) Contours depicting the variation of (a) $q^{(0)}$, (b) $q^{(0)} + Re_E q^{(1)}$ with $b$=0 and (c) $q^{(0)} + Re_E q^{(1)}$ with $b$=2 for a transverse electric field (case II). The parameters employed are $R = 0.75$, $S = 2$, $M = 2$, $\lambda = 0.25$, $Re_E = 0.2$ and $R_0 = 5$. The direction of the imposed electric field and velocity field is shown at the top.

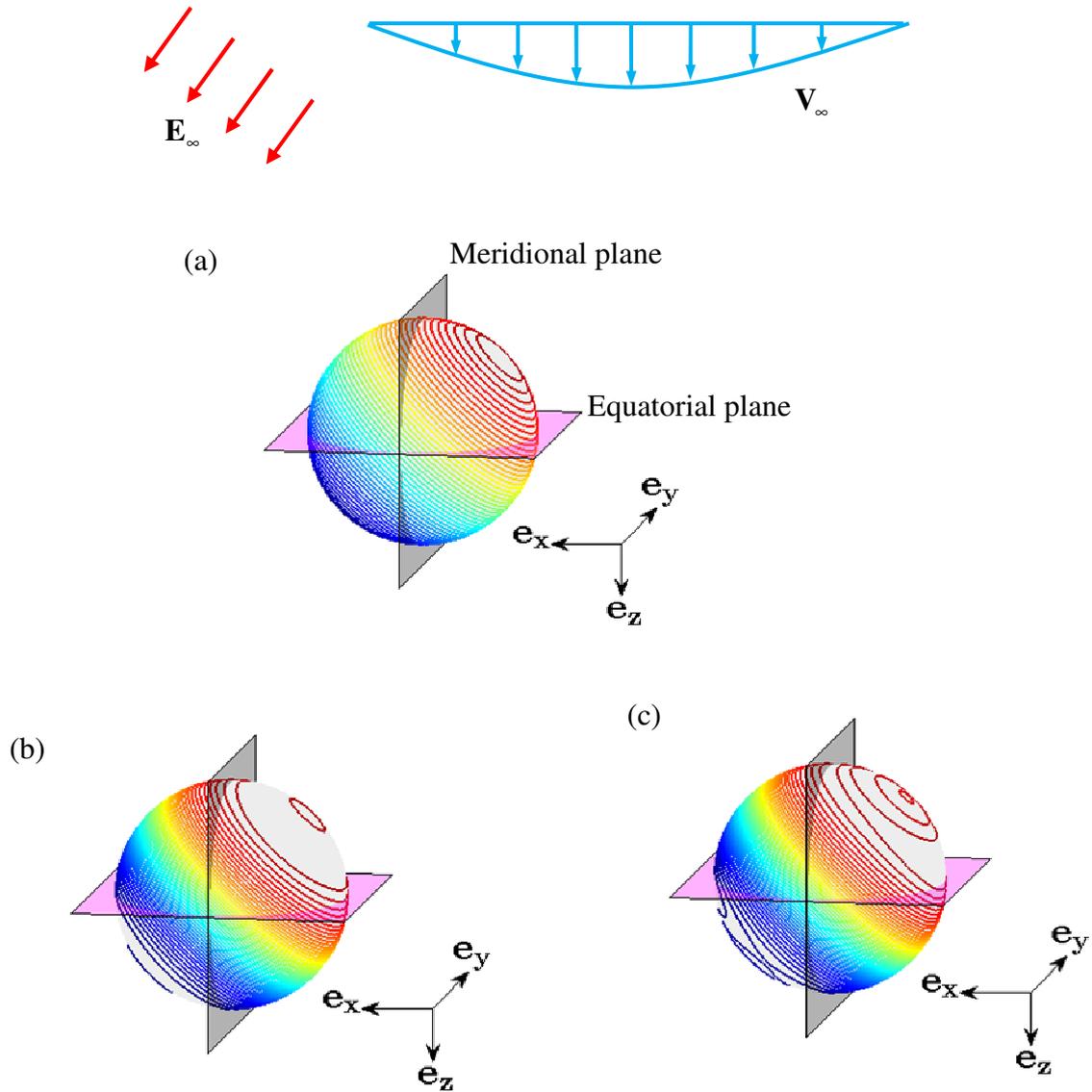

FIGURE 4. (Colour online) Contours depicting the variation of (a) $q^{(0)}$, (b) $q^{(0)} + Re_E q^{(1)}$ with $b$=0 and (c) $q^{(0)} + Re_E q^{(1)}$ with $b$=2 for a combined longitudinal and transverse electric field (case III). The parameters employed are $R = 0.75$, $S = 2$, $M = 2$, $\lambda = 0.25$, $Re_E = 0.2$ and $R_0 = 5$. The direction of the imposed electric field and velocity field is shown at the top.

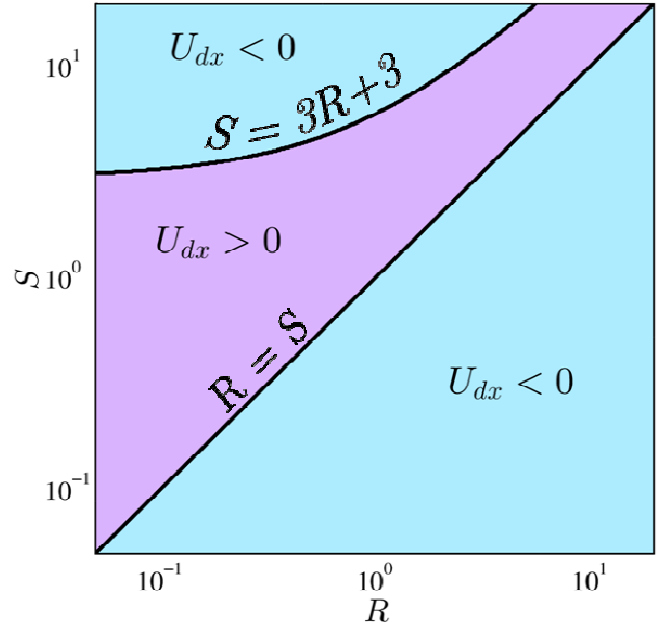

FIGURE 5. (Colour online) Different regimes of cross-stream velocity of the drop $\left(U_{dx}\right)$ in $R - S$ plane. The parameters used are $M = 2$, $\lambda = 0.25$, $Re_E = 0.2$ and $R_0 = 5$.

$$\mathbf{U}_d = \left[ 1 - \left( \frac{b}{R_0} \right)^2 - \frac{2\lambda}{R_0^2 \left( 2 + 3\lambda \right)} \right.$$
$$\left. - Re_E \left( \frac{3}{35} \frac{M \left( R - S \right) \left( 3R - S + 3 \right) \left( 43\lambda^2 + 217\lambda + 180 \right)}{R_0^2 \left( R + 2 \right)^2 \left( 3 + 2R \right) \left( 2 + 3\lambda \right)^2 \left( \lambda + 1 \right) \left( 4 + \lambda \right)} \right) \right] \mathbf{e}_z \qquad (4.12)$$
$$- Re_E \left[ \frac{3}{35} \frac{M \left( R - S \right) \left( 3R - S + 3 \right) \left( 19\lambda^2 + 91\lambda + 60 \right)}{R_0^2 \left( R + 2 \right)^2 \left( 3 + 2R \right) \left( 2 + 3\lambda \right)^2 \left( \lambda + 1 \right) \left( 4 + \lambda \right)} \right] \mathbf{e}_x .$$

In this case the drop experiences an electrical torque of the form

$$\mathbf{T}^E = - \left[ \frac{12\pi Re_E \left( R - S \right) b}{R_0^2 \left( R + 2 \right)^2} \right] \mathbf{e}_y \qquad (4.13)$$

which is balanced by the hydrodynamic torque $\mathbf{T}^H$.

An important thing to note from equation (4.12) is that the direction of the cross-stream velocity of the drop depends on the relative magnitude of the electrical conductivity ratio $\left( R \right)$ and the electrical permittivity ratio $\left( S \right)$. To show the effect of $R$ and $S$ on the cross-stream velocity of the drop, we construct a regime diagram (figure 5) in the $R - S$ plane.

From figure 5 it is clear that the two curves $R = S$ and $S = 3R + 3$ indicate the lines of zero cross-stream velocity even in the presence of non-zero $E_x$ and $E_z$.

## 5. Conclusions

In this paper, we have considered the electrohydrodynamic motion of a Newtonian, leaky dielectric, non-deforming spherical drop in a general quadratic flow in the presence of a uniform electric field acting in an arbitrary direction. Considering the effect of surface charge convection at the drop surface, we obtain the first order correction to the drop velocity using the electric Reynolds number $Re_E$ as a perturbation parameter. The presence of a background cylindrical Poiseuille flow coupled with the charge convection on the drop surface leads to a cross-stream migration of a spherical drop in the presence of simultaneous longitudinal and transverse components of the applied uniform electric field. This is in sharp contrast with a neutrally buoyant, Newtonian, non-deforming spherical drop with clean fluid-fluid interface in a cylindrical Poiseuille flow, for which the drop moves only in the axial direction (no cross-stream migration); the application of a uniform electric field having both components (longitudinal and transverse) counter-intuitively induces a cross-stream migration of the drop. Few important conclusions that can be drawn from the above results are as follows:

(1) It is apparent from the expressions given in equation (4.4) that the cross-stream migration of drop in $x$ (or $y$) direction may take place only in the presence of simultaneous non-zero $E_x$ (or $E_y$) and $E_z$ components of the applied electric field. Application of uniform electric field having a single component (only $E_x$ or $E_y$ or $E_z$) is insufficient to cause cross-stream migration of a spherical drop in the presence of a background cylindrical Poiseuille flow.

(2) Depending on the conductivity ratio $R$ and the permittivity ratio $S$, the cross-stream migration of the drop may occur towards the centreline of the Poiseuille flow or away from the centreline of the Poiseuille flow.

(3) The drop velocity in the cross-stream direction is independent of the drop offset with respect to the centreline of the Poiseuille flow, $b$.

The above calculations of the drop velocity facilitate us to acquire a preliminary idea about the controlling parameters which affect the drop velocity in the presence of uniform electric field and background flow filed. Using these, we may achieve fine-tuned control over the motion of drops in droplet-based microfluidic devices. The cross-stream migration of drops which is governed by the direction of applied electric field and the electrical properties may also be employed for sorting of drops.

## Appendix A. Derivation of the relations between the coefficients present in the solid spherical harmonics in $\mathbf{u}_\infty^{(0)}$ with the coefficients present in the solid spherical harmonics in $\mathbf{V}_\infty$ and the components of drop velocity $\mathbf{U}_d^{(0)}$

First we obtain the dot product of unit radial vector with $\mathbf{u}_\infty^{(0)}$, $\mathbf{V}_\infty$ and $\mathbf{U}_d^{(0)}$ as (Hetsroni & Haber 1970)

$$
\begin{aligned}
\mathbf{u}_\infty^{(0)} \cdot \mathbf{e}_r = \sum_{n=1}^{\infty} \sum_{m=0}^{n} \Big[ &\alpha_{n,m}^{(0)} r^{n+1} + \beta_{n,m}^{(0)} r^{n-1} + \alpha_{-n-1,m}^{(0)} r^{-n} + \beta_{-n-1}^{(0)} r^{n-2} \Big] P_{n,m} \cos m\phi \\
+ \sum_{n=1}^{\infty} \sum_{m=0}^{n} \Big[ &\hat{\alpha}_{n,m}^{(0)} r^{n+1} + \hat{\beta}_{n,m}^{(0)} r^{n-1} + \hat{\alpha}_{-n-1,m}^{(0)} r^{-n} + \hat{\beta}_{-n-1,m}^{(0)} r^{n-2} \Big] P_{n,m} \sin m\phi,
\end{aligned}
\tag{A1}
$$

$$
\begin{aligned}
\mathbf{V}_\infty \cdot \mathbf{e}_r = \sum_{n=1}^{\infty} \sum_{m=0}^{n} \Big[ &\zeta_{n,m}^{(0)} r^{n+1} + \eta_{n,m}^{(0)} r^{n-1} + \zeta_{-n-1,m}^{(0)} r^{-n} + \eta_{-n-1}^{(0)} r^{n-2} \Big] P_{n,m} \cos m\phi \\
+ \sum_{n=1}^{\infty} \sum_{m=0}^{n} \Big[ &\hat{\zeta}_{n,m}^{(0)} r^{n+1} + \hat{\eta}_{n,m}^{(0)} r^{n-1} + \hat{\zeta}_{-n-1,m}^{(0)} r^{-n} + \hat{\eta}_{-n-1,m}^{(0)} r^{n-2} \Big] P_{n,m} \sin m\phi,
\end{aligned}
\tag{A2}
$$

$$
\mathbf{U}_d^{(0)} \cdot \mathbf{e}_r = U_{dx}^{(0)} \cos\phi P_{1,1} + U_{dy}^{(0)} \sin\phi P_{1,1} + U_{dz}^{(0)} P_{1,0}.
\tag{A3}
$$

By comparing both sides of equation $\mathbf{u}_\infty^{(0)} = \mathbf{V}_\infty - \mathbf{U}_d^{(0)}$ and using equations (A1-A3) we obtain the following expressions of $\alpha_{n,m}^{(0)}, \hat{\alpha}_{n,m}^{(0)}, \beta_{n,m}^{(0)}$ and $\hat{\beta}_{n,m}^{(0)}$

$$
\begin{aligned}
\beta_{1,0}^{(0)} &= \eta_{1,0} - U_{dz}^{(0)}, \\
\beta_{1,1}^{(0)} &= \eta_{1,1} - U_{dx}^{(0)}, \\
\hat{\beta}_{1,1}^{(0)} &= \hat{\eta}_{1,1} - U_{dy}^{(0)},
\end{aligned}
\tag{A4}
$$

for all other values of $n$ and $m$ we obtain

$$
\begin{aligned}
\left\{ \alpha_{n,m}^{(0)}, \alpha_{-n-1,m}^{(0)}, \hat{\alpha}_{n,m}^{(0)}, \hat{\alpha}_{-n-1,m}^{(0)} \right\} &= \left\{ \zeta_{n,m}, \zeta_{-n-1,m}, \hat{\zeta}_{n,m}, \hat{\zeta}_{-n-1,m} \right\}, \\
\left\{ \beta_{n,m}^{(0)}, \beta_{-n-1,m}^{(0)}, \hat{\beta}_{n,m}^{(0)}, \hat{\beta}_{-n-1,m}^{(0)} \right\} &= \left\{ \eta_{n,m}, \eta_{-n-1,m}, \hat{\eta}_{n,m}, \hat{\eta}_{-n-1,m} \right\}.
\end{aligned}
\tag{A5}
$$

The coefficients $\gamma_{n,m}^{(0)}$ and $\hat{\gamma}_{n,m}^{(0)}$ can be obtained by taking $(\mathbf{r} \cdot \nabla \times)$ operation over both sides of $\mathbf{u}_\infty^{(0)} = \mathbf{V}_\infty - \mathbf{U}_d^{(0)}$. We obtain

$$
\mathbf{r} \cdot \nabla \times \mathbf{u}_\infty^{(0)} = \mathbf{r} \cdot \nabla \times \mathbf{V}_\infty - \mathbf{r} \cdot \nabla \times \mathbf{U}_d^{(0)},
\tag{A6}
$$

where $\mathbf{r} \cdot \nabla \times \mathbf{u}_\infty^{(0)}$ and $\mathbf{r} \cdot \nabla \times \mathbf{V}_\infty$ can be expressed as (Hetsroni & Haber 1970)

$$\mathbf{r} \cdot \nabla \times \mathbf{u}_\infty^{(0)} = \sum_{n=1}^{\infty} \sum_{m=0}^{n} \left[ \left\{ \gamma_{n,m}^{(0)} r^n + \gamma_{-n-1,m}^{(0)} r^{-n-1} \right\} \cos m\phi + \left\{ \hat{\gamma}_{n,m}^{(0)} r^n + \hat{\gamma}_{-n-1,m}^{(0)} r^{-n-1} \right\} \sin m\phi \right] P_{n,m},$$
(A7)

$$\mathbf{r} \cdot \nabla \times \mathbf{V}_\infty = \sum_{n=1}^{\infty} \sum_{m=0}^{n} \left[ \left\{ \varpi_{n,m}^{(0)} r^n + \varpi_{-n-1,m}^{(0)} r^{-n-1} \right\} \cos m\phi + \left\{ \hat{\varpi}_{n,m}^{(0)} r^n + \hat{\varpi}_{-n-1,m}^{(0)} r^{-n-1} \right\} \sin m\phi \right] P_{n,m}. \quad \text{(A8)}$$

Now, comparing both sides of equations (A6) and using equations (A7-A8), we obtain

$$\left\{ \gamma_{n,m}^{(0)}, \gamma_{-n-1,m}^{(0)}, \hat{\gamma}_{n,m}^{(0)}, \hat{\gamma}_{-n-1,m}^{(0)} \right\} = \left\{ \varpi_{n,m}, \varpi_{-n-1,m}, \hat{\varpi}_{n,m}, \hat{\varpi}_{-n-1,m} \right\}. \tag{A9}$$

Now, we express $\zeta_{n,m}^{(0)}, \zeta_{n,m}^{(0)}, \eta_{n,m}^{(0)}, \hat{\eta}_{n,m}^{(0)}, \varpi_{n,m}^{(0)}$ and $\hat{\varpi}_{n,m}^{(0)}$ in terms of $k_0, k_1, k_2$ and $k_3$. The unperturbed velocity at far field is a general quadratic flow filed which is in non-dimensional form is given by $\mathbf{V}_\infty = \left( k_0 + k_1 x + k_2 x^2 + k_3 y^2 \right) \mathbf{e}_z$. Substituting $\mathbf{V}_\infty$ in equations (A2) and (A8) we obtain

$$\zeta_{1,0} = \frac{k_2 + k_3}{5}, \quad \eta_{1,0} = k_0, \quad \eta_{2,1} = \frac{k_1}{3}, \quad \eta_{3,0} = -\frac{k_2 + k_3}{5},$$

$$\eta_{3,2} = \frac{k_2 - k_3}{30}, \quad \varpi_{1,1} = -k_1, \quad \varpi_{2,2} = -\frac{k_2 - k_3}{3}, \tag{A10}$$

while for all other values of $n$ and $m$ the coefficients $\zeta_{n,m}^{(0)}, \zeta_{n,m}^{(0)}, \eta_{n,m}^{(0)}, \hat{\eta}_{n,m}^{(0)}, \varpi_{n,m}^{(0)}$ and $\hat{\varpi}_{n,m}^{(0)}$ vanish.

## Appendix B. Implementation of boundary conditions to solve leading order velocity and pressure field inside and outside the drop

Towards implementing the boundary conditions (equation(3.13)), it is customary to represent the expressions $\left[ \mathbf{u}^{(0)} \cdot \mathbf{e}_r \right]_s, \left[ r \dfrac{\partial}{\partial r} \left( \mathbf{u}^{(0)} \cdot \mathbf{e}_r \right) \right]_s$ and $\left[ \mathbf{r} \cdot \nabla \times \mathbf{u}^{(0)} \right]_s$ in terms of solid spherical harmonics present in the Lamb solution in the following form (Happel & Brenner 1981)

$$\left[ \mathbf{u}^{(0)} \cdot \mathbf{e}_r \right]_s = \sum_{n=-\infty}^{\infty} \left[ \frac{n}{2N_\mu (2n+3)} p_n^{(0)} + n\Phi_n^{(0)} \right], \tag{B1}$$

$$\left[ r \frac{\partial}{\partial r} \left( \mathbf{u}^{(0)} \cdot \mathbf{e}_r \right) \right]_s = \sum_{n=-\infty}^{\infty} \left[ \frac{n(n+1)}{2N_\mu (2n+3)} p_n^{(0)} + n(n-1)\Phi_n^{(0)} \right], \tag{B2}$$

$$\left[ \mathbf{r} \cdot \nabla \times \mathbf{u}^{(0)} \right]_s = \sum_{n=-\infty}^{\infty} n(n+1) \chi_n^{(0)}. \tag{B3}$$

Now, by substituting $\mathbf{u}^{(0)} = \mathbf{u}_i^{(0)}$ and $\mathbf{u}^{(0)} = \mathbf{u}_\infty^{(0)} + \mathbf{v}_e^{(0)}$ we obtain the required expressions to satisfy the first four velocity boundary conditions given in equation (3.13). The tangential hydrodynamic stress terms $\left[ \mathbf{r} \cdot \nabla \times \left\{ \mathbf{r} \times \left( \boldsymbol{\tau}^{H(0)} \cdot \mathbf{e}_r \right) \right\} \right]_s$ and $\left[ \mathbf{r} \cdot \nabla \times \left\{ \boldsymbol{\tau}^{H(0)} \cdot \mathbf{e}_r \right\} \right]_s$ can also be represented in terms of the solid spherical harmonics present in the Lamb solution. The hydrodynamic stress components are expresses in the following form

$$\left[ \mathbf{r} \cdot \nabla \times \left( \mathbf{r} \times \left( \boldsymbol{\tau}^{H(0)} \cdot \mathbf{e}_r \right) \right) \right]_s = -N_\mu \sum_{n=-\infty}^{\infty} \left[ 2(n-1)n(n+1)\Phi_n^{(0)} + \frac{n^2(n+2)}{N_\mu(2n+3)} p_n^{(0)} \right], \quad \text{(B4)}$$

$$\left[ \mathbf{r} \cdot \nabla \times \left( \boldsymbol{\tau}^{H(0)} \cdot \mathbf{e}_r \right) \right]_s = N_\mu \sum_{n=-\infty}^{\infty} (n-1)n(n+1)\chi_n^{(0)}. \quad \text{(B5)}$$

From the Maxwell stress tensor, we obtain the radial traction vector $\boldsymbol{\tau}_i^{E(0)} \cdot \mathbf{e}_r$ and $\boldsymbol{\tau}_e^{E(0)} \cdot \mathbf{e}_r$ in the following form

$$\boldsymbol{\tau}_i^{E(0)} \cdot \mathbf{e}_r = S \begin{bmatrix} \left( E_{i,r}^{(0)} \right)^2 - \frac{1}{2} \left| \mathbf{E}_i^{(0)} \right|^2 \\ E_{i,r}^{(0)} E_{i,\theta}^{(0)} \\ E_{i,r}^{(0)} E_{i,\phi}^{(0)} \end{bmatrix}, \quad \boldsymbol{\tau}_e^{E(0)} \cdot \mathbf{e}_r = \begin{bmatrix} \left( E_{e,r}^{(0)} \right)^2 - \frac{1}{2} \left| \mathbf{E}_e^{(0)} \right|^2 \\ E_{e,r}^{(0)} E_{e,\theta}^{(0)} \\ E_{e,r}^{(0)} E_{e,\phi}^{(0)} \end{bmatrix}. \quad \text{(B6)}$$

Now, the electrical stress components are obtained in terms of spherical surface harmonics as

$$\begin{aligned}
\left[ \mathbf{r} \cdot \nabla \times \left\{ \mathbf{r} \times \left( \boldsymbol{\tau}_i^{E(0)} \cdot \mathbf{e}_r \right) \right\} \right]_s &= \sum_{n=0}^{\infty} \left( g_{n,m}^{i(0)} \cos m\phi + \hat{g}_{n,m}^{i(0)} \sin m\phi \right) P_{n,m}, \\
\left[ \mathbf{r} \cdot \nabla \times \left\{ \mathbf{r} \times \left( \boldsymbol{\tau}_i^{E(0)} \cdot \mathbf{e}_r \right) \right\} \right]_s &= \sum_{n=0}^{\infty} \left( g_{n,m}^{e(0)} \cos m\phi + \hat{g}_{n,m}^{e(0)} \sin m\phi \right) P_{n,m}, \\
\left[ \mathbf{r} \cdot \nabla \times \left( \boldsymbol{\tau}_i^{E(0)} \cdot \mathbf{e}_r \right) \right]_s &= \sum_{n=0}^{\infty} \left( h_{n,m}^{i(0)} \cos m\phi + \hat{h}_{n,m}^{i(0)} \sin m\phi \right) P_{n,m}, \\
\left[ \mathbf{r} \cdot \nabla \times \left( \boldsymbol{\tau}_e^{E(0)} \cdot \mathbf{e}_r \right) \right]_s &= \sum_{n=0}^{\infty} \left( h_{n,m}^{e(0)} \cos m\phi + \hat{h}_{n,m}^{e(0)} \sin m\phi \right) P_{n,m}.
\end{aligned} \quad \text{(B7)}$$

The coefficients $g_{n,m}^{i(0)}, \hat{g}_{n,m}^{i(0)}, h_{n,m}^{i(0)}, \hat{h}_{n,m}^{i(0)}, g_{n,m}^{e(0)}, \hat{g}_{n,m}^{e(0)}, h_{n,m}^{e(0)}$ and $\hat{h}_{n,m}^{e(0)}$ are obtained after obtaining the leading order electrical potential. The non-zero coefficients are obtained as:

$$g_{2,0}^{i(0)} = \frac{9\left(E_y^2 + E_x^2 - 2E_z^2\right)S}{\left(R+2\right)^2}, \, g_{2,1}^{i(0)} = -\frac{18SE_zE_x}{\left(R+2\right)^2}, \, g_{2,2}^{i(0)} = -\frac{9}{2}\frac{\left(E_x^2 - E_y^2\right)S}{\left(R+2\right)^2},$$

$$\hat{g}_{2,1}^{i(0)} = \frac{-18E_yE_zS}{\left(R+2\right)^2}, \, \hat{g}_{2,2}^{i(0)} = -\frac{9E_xE_yS}{\left(R+2\right)^2}, \, g_{2,0}^{e(0)} = \frac{9\left(E_x^2 + E_y^2 - 2E_z^2\right)R}{\left(R+2\right)^2},$$

$$g_{2,1}^{e(0)} = \frac{-18\,R\,E_zE_x}{\left(R+2\right)^2}, \, g_{2,2}^{e(0)} = -\frac{9}{2}\frac{\left(-E_y^2 + E_x^2\right)R}{\left(R+2\right)^2}, \, \hat{g}_{2,1}^{e(0)} = \frac{-18E_yE_zR}{\left(R+2\right)^2},$$

$$\hat{g}_{2,2}^{e(0)} = -\frac{9E_xE_yR}{\left(R+2\right)^2}.$$

(B8)

## Appendix C. Velocity and pressure field inside and outside the drop at the leading order

The velocity field and the pressure field inside the drop is obtained as

$$\mathbf{u}_i^{(0)} = \left[\nabla\times\left(\mathbf{r}\chi_1^{(0)} + \mathbf{r}\chi_2^{(0)}\right) + \nabla\left(\Phi_1^{(0)} + \Phi_2^{(0)}\right) + r^2\left(\frac{1}{5\lambda}\nabla p_1^{(0)} + \frac{5}{42\lambda}\nabla p_2^{(0)} + \frac{1}{12\lambda}\nabla p_3^{(0)}\right)\right.$$
$$\left. -\mathbf{r}\left(\frac{1}{10\lambda}\,p_1^{(0)} + \frac{2}{21\lambda}\,p_2^{(0)} + \frac{1}{12\lambda}\,p_3^{(0)}\right)\right],$$

(C1)

$$p_i^{(0)} = p_1^{(0)} + p_2^{(0)} + p_3^{(0)}.$$

The solid harmonics present in equation (C1) are obtained in the following form

$$\chi_1^{(0)} = r\hat{C}_{1,1}^{(0)}\sin\phi P_{1,1},$$
$$\chi_2^{(0)} = r^2\hat{C}_{2,2}^{(0)}\sin2\phi P_{2,2},$$
$$\Phi_1^{(0)} = rB_{1,0}^{(0)}P_{1,0},$$
$$\Phi_2^{(0)} = r^2\left[B_{2,0}^{(0)}P_{2,0} + \left(B_{2,1}^{(0)}\cos\phi + \hat{B}_{2,1}^{(0)}\sin\phi\right)P_{2,1} + \left(B_{2,2}^{(0)}\cos2\phi + \hat{B}_{2,2}^{(0)}\sin2\phi\right)P_{2,2}\right],$$
$$\Phi_3^{(0)} = r^3\left[B_{3,0}^{(0)}P_{3,0} + \left(B_{3,2}^{(0)}\cos2\phi\right)P_{3,2}\right],$$
$$p_1^{(0)} = \lambda rA_{1,0}^{(0)}P_{1,0},$$
$$p_2^{(0)} = \lambda r^2\left[A_{2,0}^{(0)}P_{2,0} + \left(A_{2,1}^{(0)}\cos\phi + \hat{A}_{2,1}^{(0)}\sin\phi\right)P_{2,1} + \left(A_{2,2}^{(0)}\cos2\phi + \hat{A}_{2,2}^{(0)}\sin2\phi\right)P_{2,2}\right],$$
$$p_3^{(0)} = \lambda r^3\left[A_{3,0}^{(0)}P_{3,0} + \left(A_{3,2}^{(0)}\cos2\phi\right)P_{3,2}\right].$$

(C2)

We obtain the coefficients $A^{(0)}, \hat{A}^{(0)}, B^{(0)}, \hat{B}^{(0)}, C^{(0)}$ and $\hat{C}^{(0)}$ present in equation (C2) by using equation (3.14) as

$$A_{1,0}^{(0)} = \frac{10(k_2 + k_3)}{3\lambda + 2}, \quad B_{1,0}^{(0)} = -\frac{A_{1,0}^{(0)}}{10}, \quad A_{2,0}^{(0)} = -\frac{63}{10}\frac{M\left(E_y^2 + E_x^2 - 2E_z^2\right)(R-S)}{(R+2)^2(\lambda+1)},$$

$$B_{2,0}^{(0)} = -\frac{A_{2,0}^{(0)}}{14}, \quad A_{2,1}^{(0)} = \frac{7}{10}\frac{5k_1(R+2)^2 + 18ME_zE_x(R-S)}{(R+2)^2(\lambda+1)}, \quad B_{2,1}^{(0)} = -\frac{A_{2,1}^{(0)}}{14},$$

$$A_{2,2}^{(0)} = \frac{63}{20}\frac{M\left(E_x^2 - E_y^2\right)(R-S)}{(R+2)^2(\lambda+1)}, \quad B_{2,2}^{(0)} = -\frac{A_{2,2}^{(0)}}{14}, \quad A_{3,0}^{(0)} = -\frac{3(k_2 + k_3)}{\lambda+1}, \quad B_{3,0}^{(0)} = -\frac{A_{3,0}^{(0)}}{18},$$

$$A_{3,2}^{(0)} = \frac{1}{2}\frac{k_2 - k_3}{\lambda+1}, \quad B_{3,2}^{(0)} = -\frac{A_{3,2}^{(0)}}{18}, \quad \hat{A}_{2,1}^{(0)} = \frac{63}{5}\frac{ME_yE_z(R-S)}{(R+2)^2(\lambda+1)}, \quad \hat{B}_{2,1}^{(0)} = -\frac{9}{10}\frac{ME_yE_z(R-S)}{(R+2)^2(\lambda+1)},$$

$$\hat{A}_{2,2}^{(0)} = \frac{63}{10}\frac{ME_xE_y(R-S)}{(R+2)^2(\lambda+1)}, \quad \hat{B}_{2,2}^{(0)} = -\frac{9}{20}\frac{ME_xE_y(-S+R)}{(R+2)^2(\lambda+1)}, \quad \hat{C}_{1,1}^{(0)} = -\frac{k_1}{2}, \quad \hat{C}_{2,2}^{(0)} = -\frac{5}{18}\frac{k_2 - k_3}{4+\lambda}.$$

$$\text{(C3)}$$

The velocity field and the pressure field outside the drop is obtained as

$$
\begin{aligned}
\mathbf{u}_e^{(0)} &= \mathbf{u}_\infty^{(0)} + \mathbf{v}_e^{(0)} \\
&= \mathbf{u}_\infty^{(0)} + \sum_{n=1}^{\infty}\left[\nabla\times\left(\mathbf{r}\chi_{-2}^{(0)} + \mathbf{r}\chi_{-3}^{(0)}\right) + \nabla\left(\Phi_{-2}^{(0)} + \Phi_{-3}^{(0)} + \Phi_{-4}^{(0)}\right) - \frac{1}{30}r^2\nabla p_{-4}^{(0)} \right. \\
&\qquad\qquad \left. + \mathbf{r}\left(\frac{1}{2}p_{-3}^{(0)} + \frac{4}{15}p_{-4}^{(0)}\right)\right], \\
p_e^{(0)} &= p_{-3}^{(0)} + p_{-4}^{(0)},
\end{aligned}
\tag{C4}
$$

The solid harmonics present in equation (C4) are obtained in the following form

$$\chi_{-2}^{(0)} = \frac{1}{r^2}\hat{C}_{-2,1}^{(0)}\sin\phi\, P_{1,1},$$

$$\chi_{-3}^{(0)} = \frac{1}{r^3}\hat{C}_{-3,2}^{(0)}\sin 2\phi\, P_{2,2},$$

$$\Phi_{-2}^{(0)} = \frac{1}{r^2}B_{-2,0}^{(0)}P_{1,0},$$

$$\Phi_{-3}^{(0)} = \frac{1}{r^3}\left[B_{-3,0}^{(0)}P_{2,0} + \left(B_{-3,1}^{(0)}\cos\phi + \hat{B}_{-3,1}^{(0)}\sin\phi\right)P_{2,1} + \left(B_{-3,2}^{(0)}\cos 2\phi + \hat{B}_{-3,2}^{(0)}\sin 2\phi\right)P_{2,2}\right], \quad \text{(C5)}$$

$$\Phi_{-4}^{(0)} = \frac{1}{r^4}\left[B_{-4,0}^{(0)}P_{3,0} + \left(B_{-4,2}^{(0)}\cos 2\phi\right)P_{3,2}\right],$$

$$p_{-3}^{(0)} = \frac{1}{r^3}\left[A_{-3,0}^{(0)}P_{2,0} + \left(A_{-3,1}^{(0)}\cos\phi + \hat{A}_{-3,1}^{(0)}\sin\phi\right)P_{2,1} + \left(A_{-3,2}^{(0)}\cos 2\phi + \hat{A}_{-3,2}^{(0)}\sin 2\phi\right)P_{2,2}\right],$$

$$p_{-4}^{(0)} = \frac{1}{r^4}\left[A_{-4,0}^{(0)}P_{3,0} + \left(A_{-4,2}^{(0)}\cos 2\phi\right)P_{3,2}\right].$$

We obtain the coefficients $A^{(0)}, \hat{A}^{(0)}, B^{(0)}, \hat{B}^{(0)}, C^{(0)}$ and $\hat{C}^{(0)}$ present in equation (C5) by using equation (3.14) as

$$B_{-2,0}^{(0)} = -\frac{1}{5}\frac{(\lambda-1)(k_2+k_3)}{3\lambda+2}, \quad A_{-3}^{(0)} = -\frac{9}{5}\frac{M\left(E_y^2+E_x^2-2E_z^2\right)(R-S)}{(R+2)^2(\lambda+1)},$$

$$B_{-3,0}^{(0)} = -\frac{3}{10}\frac{M\left(E_y^2+E_x^2-2E_z^2\right)(R-S)}{(R+2)^2(\lambda+1)}, \quad A_{-3,1}^{(0)} = -\frac{1}{15}\frac{(R+2)^2(25\lambda+10)k_1-54ME_xE_z(R-S)}{(R+2)^2(\lambda+1)},$$

$$B_{-3,1}^{(0)} = -\frac{1}{30}\frac{5\lambda k_1(R+2)^2-18ME_xE_z(R-S)}{(R+2)^2(\lambda+1)}, \quad A_{-3,2}^{(0)} = \frac{9}{10}\frac{W\left(E_x^2-E_y^2\right)(-S+R)}{(R+2)^2(\lambda+1)},$$

<div align="right">(C6)</div>

and

$$B_{-3,2}^{(0)} = \frac{3}{20}\frac{W\left(E_x^2-E_y^2\right)(R-S)}{(R+2)^2(\lambda+1)}, \quad A_{-4,0}^{(0)} = \frac{1}{4}\frac{(7\lambda+2)(k_2+k_3)}{\lambda+1}, \quad B_{-4,0}^{(0)} = \frac{1}{8}\frac{\lambda(k_2+k_3)}{\lambda+1},$$

$$A_{-4,2}^{(0)} = -\frac{1}{24}\frac{(7\lambda+2)(k_2-k_3)}{\lambda+1}, \quad B_{-4,2}^{(0)} = -\frac{1}{48}\frac{\lambda(k_2-k_3)}{\lambda+1}, \quad \hat{A}_{-3,1}^{(0)} = \frac{18}{5}\frac{ME_yE_z(R-S)}{(R+2)^2(\lambda+1)},$$

$$\hat{B}_{-3,1}^{(0)} = \frac{3}{5}\frac{ME_yE_z(R-S)}{(R+2)^2(\lambda+1)}, \quad \hat{A}_{-3,2}^{(0)} = \frac{9}{5}\frac{ME_xE_y(R-S)}{(R+2)^2(\lambda+1)}, \quad \hat{B}_{-3,2}^{(0)} = \frac{3}{10}\frac{ME_xE_y(R-S)}{(R+2)^2(\lambda+1)},$$

$$\hat{C}_{-3,2}^{(0)} = \frac{1}{18}\frac{(\lambda-1)(k_2-k_3)}{4+\lambda}.$$

<div align="right">(C7)</div>

## Appendix D. Complete expressions of the coefficients $Z_{n,m}$ present at the right hand side of the charge conservation equation at $O\left(Re_E\right)$

The complete expressions of the non-zero coefficients $Z_{n,m}$ are obtained as

$$Z_{1,0} = \frac{3}{50} \frac{(R-S)}{(R+2)^3 (\lambda+1)} \left[ \left\{ E_z (R+2)^2 (25\lambda+10) \right\} k_1 + 36M (R-S) E_z \left( E_x^2 + E_y^2 + E_z^2 \right) \right],$$

$$Z_{1,1} = -\frac{3}{50} \frac{(R-S)}{(R+2)^3 (\lambda+1)} \left[ \left\{ E_z (R+2)^2 (25\lambda+40) \right\} k_1 - 36M (R-S) E_x \left( E_x^2 + E_y^2 + E_z^2 \right) \right],$$

$$Z_{2,0} = \frac{6}{7} \frac{E_z (k_2 + k_3)(R-S)(4\lambda-5)}{(3\lambda+2)(\lambda+1)(R+2)},$$

$$Z_{2,1} = \frac{2}{7} \frac{E_x (R-S) \left[ k_2 \left( 36\lambda^2 + 119\lambda + 75 \right) + k_3 \left( 9\lambda^2 + 14\lambda + 25 \right) \right]}{(3\lambda+2)(\lambda+1)(4+\lambda)(R+2)},$$

$$Z_{2,2} = -\frac{5}{7} \frac{E_z (k_2 - k_3)(R-S)(1+2\lambda)}{(4+\lambda)(\lambda+1)(R+2)},$$

$$Z_{3,0} = -\frac{12}{25} \frac{(R-S) \left[ 5E_x (R+2)^2 k_1 + 9M (R-S) \left( 3E_x^2 + 3E_y^2 - 2E_z^2 \right) E_z \right]}{(\lambda+1)(R+2)^3},$$

$$Z_{3,1} = \frac{2}{25} \frac{(R-S) \left[ 20E_x (R+2)^2 k_1 + 27M (R-S) E_x \left( -E_x^2 - E_y^2 + 4E_z^2 \right) \right]}{(\lambda+1)(R+2)^3},$$

$$Z_{3,2} = \frac{2}{25} \frac{(R-S) \left[ 5E_x (R+2)^2 k_1 + 27M (R-S) E_x \left( E_x^2 - E_y^2 \right) \right]}{(\lambda+1)(R+2)^3},$$

$$Z_{3,3} = \frac{9}{25} \frac{M (R-S)^2 \left( E_x^2 - 3E_y^2 \right) E_x}{(R+2)^3 (\lambda+1)}, \quad Z_{4,0} = -\frac{15}{7} \frac{(k_2 + k_3)(R-S) E_z}{(\lambda+1)(R+2)},$$

$$Z_{4,1} = -\frac{15}{56} \frac{(R-S)(3k_2 + k_3) E_x}{(\lambda+1)(R+2)}, \quad Z_{4,2} = \frac{5}{28} \frac{(R-S)(k_2 - k_3) E_z}{(\lambda+1)(R+2)},$$

$$Z_{4,3} = \frac{5}{112} \frac{(R-S)(k_2 - k_3) E_x}{(\lambda+1)(R+2)}, \quad \hat{Z}_{1,1} = \frac{54}{25} \frac{W (R-S)^2 \left( E_x^2 + E_y^2 + E_z^2 \right) E_y}{(R+2)^3 (\lambda+1)},$$

$$\hat{Z}_{2,1} = -\frac{2}{7} \frac{(R-S) \left[ 9 (k_2 - 4k_3) \lambda^2 + (-119k_3 - 14k_2) \lambda - 25k_2 - 75k_3 \right] E_y}{(\lambda+1)(3\lambda+2)(4+\lambda)(R+2)},$$

$$\hat{Z}_{3,1} = -\frac{54}{25} \frac{M (R-S)^2 \left( E_x^2 + E_y^2 - 4E_z^2 \right) E_y}{(R+2)^3 (\lambda+1)},$$

$$\hat{Z}_{3,2} = \frac{2}{25} \frac{(R-S) \left[ 5 (R+2)^2 k_1 + 54M E_x E_z (R-S) \right] E_y}{(R+2)^3 (\lambda+1)},$$

$$(D1)$$

and

$$\hat{Z}_{3,3} = \frac{9}{25} \frac{M(R-S)^2 \left(3E_x^2 - E_y^2\right) E_y}{(R+2)^3 (\lambda+1)}, \quad \hat{Z}_{4,1} = -\frac{15}{56} \frac{(R-S)(k_2 + 3k_3) E_y}{(\lambda+1)(R+2)},$$

$$\hat{Z}_{4,3} = \frac{5}{112} \frac{(R-S)(k_2 - k_3) E_y}{(\lambda+1)(R+2)}.$$

<div align="right">(D2)</div>

## Appendix E. The electric potential at $O\left(Re_E\right)$

The electric potential inside and outside the drop at $O\left(Re_E\right)$ is obtained as

$$\psi_i^{(1)} = r\left[ a_{1,0}^{(1)} P_{1,0} + \left(a_{1,1}^{(1)} \cos\phi + \hat{a}_{1,1}^{(1)} \sin\phi\right) P_{1,1} \right] + r^2 \left[ a_{2,0}^{(1)} P_{2,0} + \left(a_{2,1}^{(1)} \cos\phi + \hat{a}_{2,1}^{(1)} \sin\phi\right) P_{2,1} + a_{2,2}^{(1)} \cos 2\phi P_{2,2} \right]$$
$$+ r^3 \left[ a_{3,0}^{(1)} P_{3,0} + \left(a_{3,1}^{(1)} \cos\phi + \hat{a}_{3,1}^{(1)} \sin\phi\right) P_{3,1} + \left(a_{3,2}^{(1)} \cos 2\phi + \hat{a}_{3,2}^{(1)} \sin 2\phi\right) P_{3,2} \right.$$
$$+ \left.\left(a_{3,3}^{(1)} \cos 3\phi + \hat{a}_{3,3}^{(1)} \sin 3\phi\right) P_{3,3} \right] + r^4 \left[ a_{4,0}^{(1)} P_{4,0} + \left(a_{4,1}^{(1)} \cos\phi + \hat{a}_{4,1}^{(1)} \sin\phi\right) P_{4,1} + a_{4,2}^{(1)} \cos 2\phi P_{4,2} \right.$$
$$+ \left.\left(a_{4,3}^{(1)} \cos 3\phi + \hat{a}_{4,3}^{(1)} \sin 3\phi\right) P_{4,3} \right],$$

<div align="right">(E1)</div>

$$\psi_e^{(1)} = \frac{1}{r^2}\left[ b_{-2,0}^{(1)} P_{1,0} + \left(b_{-2,1}^{(1)} \cos\phi + \hat{b}_{-2,1}^{(1)} \sin\phi\right) P_{1,1} + b_{-3,2}^{(1)} \cos 2\phi P_{2,2} \right] + \frac{1}{r^3}\left[ b_{-3,0}^{(1)} P_{2,0} + \left(b_{-3,1}^{(1)} \cos\phi + \hat{b}_{-3,1}^{(1)} \sin\phi\right) P_{2,1} \right]$$
$$+ \frac{1}{r^4}\left[ b_{-4,0}^{(1)} P_{3,0} + \left(b_{-4,1}^{(1)} \cos\phi + \hat{b}_{-4,1}^{(1)} \sin\phi\right) P_{3,1} + \left(b_{-4,2}^{(1)} \cos 2\phi + \hat{b}_{-4,2}^{(1)} \sin 2\phi\right) P_{3,2} \right.$$
$$+ \left.\left(b_{-4,3}^{(1)} \cos 3\phi + \hat{b}_{-4,3}^{(1)} \sin 3\phi\right) P_{3,3} \right] + \frac{1}{r^5}\left[ b_{-5,0}^{(1)} P_{4,0} + \left(b_{-5,1}^{(1)} \cos\phi + \hat{b}_{-5,1}^{(1)} \sin\phi\right) P_{4,1} + b_{-5,2}^{(1)} \cos 2\phi P_{4,2} \right.$$
$$+ \left.\left(b_{-5,3}^{(1)} \cos 3\phi + \hat{b}_{-5,3}^{(1)} \sin 3\phi\right) P_{4,3} \right].$$

<div align="right">(E2)</div>

The non-zero coefficients present in the above expressions are obtained in terms of $Z_{n,m}$ in the following form

$$a_{1,0}^{(1)} = \frac{Z_{1,0}}{2+R}, \quad a_{1,1}^{(1)} = \frac{Z_{1,1}}{2+R}, \quad a_{2,0}^{(1)} = \frac{Z_{2,0}}{3+2R}, \quad a_{2,1}^{(1)} = \frac{Z_{2,1}}{3+2R}, a_{2,2}^{(1)} = \frac{Z_{2,2}}{3+2R}, \quad a_{3,0}^{(1)} = \frac{Z_{3,0}}{4+3R},$$

$$a_{3,1}^{(1)} = \frac{Z_{3,1}}{4+3R}, \quad a_{3,2}^{(1)} = \frac{Z_{3,2}}{4+3R}, \quad a_{3,3}^{(1)} = \frac{Z_{3,3}}{4+3R}, a_{4,0}^{(1)} = \frac{Z_{4,0}}{5+4R}, \quad a_{4,1}^{(1)} = \frac{Z_{4,1}}{5+4R}, a_{4,2}^{(1)} = \frac{Z_{4,2}}{5+4R},$$

$$a_{4,3}^{(1)} = \frac{Z_{4,3}}{5+4R}, \quad \hat{a}_{1,1}^{(1)} = \frac{\hat{Z}_{1,1}}{2+R}, \quad \hat{a}_{2,1}^{(1)} = \frac{\hat{Z}_{2,1}}{3+2R}, \quad \hat{a}_{3,1}^{(1)} = \frac{\hat{Z}_{3,1}}{4+3R}, \quad \hat{a}_{3,2}^{(1)} = \frac{\hat{Z}_{3,2}}{4+3R}, \quad \hat{a}_{3,3}^{(1)} = \frac{\hat{Z}_{3,3}}{4+3R},$$

$$\hat{a}_{4,1}^{(1)} = \frac{\hat{Z}_{4,1}}{5+4R}, \quad \hat{a}_{4,3}^{(1)} = \frac{\hat{Z}_{4,3}}{5+4R},$$

$$b_{-2,0}^{(1)} = a_{1,0}^{(1)}, \quad b_{-2,1}^{(1)} = a_{1,1}^{(1)}, \quad b_{-3,0}^{(1)} = a_{2,0}^{(1)}, \quad b_{-3,1}^{(1)} = a_{2,1}^{(1)}, \quad b_{-3,2}^{(1)} = a_{2,2}^{(1)}, b_{-4,0}^{(1)} = a_{3,0}^{(1)}, \quad b_{-4,1}^{(1)} = a_{3,1}^{(1)},$$

$$b_{-4,2}^{(1)} = a_{3,2}^{(1)}, \quad b_{-4,3}^{(1)} = a_{3,3}^{(1)}, \quad b_{-5,0}^{(1)} = a_{4,0}^{(1)}, \quad b_{-5,1}^{(1)} = a_{4,1}^{(1)}, b_{-5,2}^{(1)} = a_{4,2}^{(1)}, \quad b_{-5,3}^{(1)} = a_{4,3}^{(1)}, \hat{b}_{-2,1}^{(1)} = \hat{a}_{1,1}^{(1)},$$

$$\hat{b}_{-3,1}^{(1)} = \hat{a}_{2,1}^{(1)}, \quad \hat{b}_{-4,1}^{(1)} = \hat{a}_{3,1}^{(1)}, \quad \hat{b}_{-4,2}^{(1)} = \hat{a}_{3,2}^{(1)}, \quad \hat{b}_{-4,3}^{(1)} = \hat{a}_{3,3}^{(1)}, \hat{b}_{-5,1}^{(1)} = \hat{a}_{4,1}^{(1)}, \quad \hat{b}_{-5,3}^{(1)} = \hat{a}_{4,3}^{(1)}.$$

$$(E3)$$

## Appendix F. The non-zero solid harmonics present in velocity and pressure field of $O(Re_E)$

Towards implementing the boundary condition for velocity field given in equation (2.12), first we express $\left[ \mathbf{u}^{(1)} \cdot \mathbf{e}_r \right]_s$, $\left[ r \frac{\partial}{\partial r} \left( \mathbf{u}^{(1)} \cdot \mathbf{e}_r \right) \right]_s$ and $\left[ \mathbf{r} \cdot \nabla \times \mathbf{u}^{(1)} \right]_s$ in terms of the solid spherical harmonics present in Lamb solution in the following form (Happel & Brenner 1981)

$$\left[ \mathbf{u}^{(1)} \cdot \mathbf{e}_r \right]_s = \sum_{n=-\infty}^{\infty} \left[ \frac{n}{2N_\mu (2n+3)} p_n^{(1)} + n\Phi_n^{(1)} \right], \tag{F1}$$

$$\left[ r \frac{\partial}{\partial r} \left( \mathbf{u}^{(1)} \cdot \mathbf{e}_r \right) \right]_s = \sum_{n=-\infty}^{\infty} \left[ \frac{n(n+1)}{2N_\mu (2n+3)} p_n^{(1)} + n(n-1)\Phi_n^{(1)} \right],$$

$$\left[ \mathbf{r} \cdot \nabla \times \mathbf{u}^{(1)} \right]_s = \sum_{n=-\infty}^{\infty} n(n+1) \chi_n^{(1)}. \tag{F2}$$

To determine the stress boundary condition, we evaluate $\left[ \mathbf{r} \cdot \nabla \times \left\{ \mathbf{r} \times \left( \left( \boldsymbol{\tau}^{H(1)} + M\boldsymbol{\tau}^{E(1)} \right) \cdot \mathbf{e}_r \right) \right\} \right]_s$ and $\left[ \mathbf{r} \cdot \nabla \times \left\{ \left( \boldsymbol{\tau}^{H(1)} + M\boldsymbol{\tau}^{E(1)} \right) \cdot \mathbf{e}_r \right\} \right]_s$. The hydrodynamic stress components are expressed as

$$\left[ \mathbf{r} \cdot \nabla \times \left( \mathbf{r} \times \left( \boldsymbol{\tau}^{H(1)} \cdot \mathbf{e}_r \right) \right) \right]_s = -N_\mu \sum_{n=-\infty}^{\infty} \left[ 2(n-1)n(n+1)\Phi_n^{(1)} + \frac{n^2(n+2)}{N_\mu (2n+3)} p_n^{(1)} \right], \tag{F3}$$

$$\left[ \mathbf{r} \cdot \nabla \times \left( \boldsymbol{\tau}^{H(1)} \cdot \mathbf{e}_r \right) \right]_s = N_\mu \sum_{n=-\infty}^{\infty} (n-1) n (n+1) \chi_n^{(1)}. \tag{F4}$$

The normal component of the electrical traction vector $\boldsymbol{\tau}_i^{E(1)} \cdot \mathbf{e}_r$ and $\boldsymbol{\tau}_e^{E(1)} \cdot \mathbf{e}_r$ at $O\left( Re_E \right)$ are expressed as

$$\boldsymbol{\tau}_i^{E(1)} \cdot \mathbf{e}_r = S \begin{bmatrix} E_{i,r}^{(0)} E_{i,r}^{(1)} - \dfrac{1}{2} \left( 2 E_{i,r}^{(0)} E_{i,r}^{(1)} + 2 E_{i,\theta}^{(0)} E_{i,\theta}^{(1)} + 2 E_{i,\phi}^{(0)} E_{i,\phi}^{(1)} \right) \\ \left( E_{i,r}^{(0)} E_{i,\theta}^{(1)} + E_{i,r}^{(1)} E_{i,\theta}^{(0)} \right) \\ \left( E_{i,r}^{(0)} E_{i,\phi}^{(1)} + E_{i,r}^{(1)} E_{i,\phi}^{(0)} \right) \end{bmatrix}, \tag{F5}$$

$$\boldsymbol{\tau}_e^{E(1)} \cdot \mathbf{e}_r = \begin{bmatrix} E_{e,r}^{(0)} E_{e,r}^{(1)} - \dfrac{1}{2} \left( 2 E_{e,r}^{(0)} E_{e,r}^{(1)} + 2 E_{e,\theta}^{(0)} E_{e,\theta}^{(1)} + 2 E_{e,\phi}^{(0)} E_{e,\phi}^{(1)} \right) \\ \left( E_{e,r}^{(0)} E_{e,\theta}^{(1)} + E_{e,r}^{(1)} E_{e,\theta}^{(0)} \right) \\ \left( E_{e,r}^{(0)} E_{e,\phi}^{(1)} + E_{e,r}^{(1)} E_{e,\phi}^{(0)} \right) \end{bmatrix}. \tag{F6}$$

The electrical stress components at the drop surface can be expressed as linear combination of surface harmonics of the form

$$\left[ \mathbf{r} \cdot \nabla \times \left\{ \mathbf{r} \times \left( \boldsymbol{\tau}_i^{E(1)} \cdot \mathbf{e}_r \right) \right\} \right]_s = \sum_{n=0}^{\infty} \left( g_{n,m}^{i(1)} \cos m\phi + \hat{g}_{n,m}^{i(1)} \sin m\phi \right) P_{n,m},$$

$$\left[ \mathbf{r} \cdot \nabla \times \left\{ \mathbf{r} \times \left( \boldsymbol{\tau}_e^{E(1)} \cdot \mathbf{e}_r \right) \right\} \right]_s = \sum_{n=0}^{\infty} \left( g_{n,m}^{e(1)} \cos m\phi + \hat{g}_{n,m}^{e(1)} \sin m\phi \right) P_{n,m}, \tag{F7}$$

$$\left[ \mathbf{r} \cdot \nabla \times \left( \boldsymbol{\tau}_i^{E(1)} \cdot \mathbf{e}_r \right) \right]_s = \sum_{n=0}^{\infty} \left( h_{n,m}^{e(1)} \cos m\phi + \hat{h}_{n,m}^{e(1)} \sin m\phi \right) P_{n,m},$$

$$\left[ \mathbf{r} \cdot \nabla \times \left( \boldsymbol{\tau}_e^{E(1)} \cdot \mathbf{e}_r \right) \right]_s = \sum_{n=0}^{\infty} \left( h_{n,m}^{e(1)} \cos m\phi + \hat{h}_{n,m}^{e(1)} \sin m\phi \right) P_{n,m}, \tag{F8}$$

where the coefficients $g_{n,m}^{i(1)}$, $\hat{g}_{n,m}^{i(1)}$, $h_{n,m}^{i(1)}$, $\hat{h}_{n,m}^{i(1)}$, $g_{n,m}^{e(1)}$, $\hat{g}_{n,m}^{e(1)}$, $h_{n,m}^{e(1)}$ and $\hat{h}_{n,m}^{e(1)}$ are obtained after solving electrical potential distribution at the first order of approximation and the non-zero coefficients are listed in table 2. The complete expressions of the coefficients of the terms given in table 2 are too large to present and we only provide those coefficients which are necessary for obtaining the drop velocity in the following form

| $g_{n,m}^{i(1)}$ and $\hat{g}_{n,m}^{i(1)}$ | $g_{n,m}^{e(1)}$ and $\hat{g}_{n,m}^{e(1)}$ | $h_{n,m}^{i(1)}$ and $\hat{h}_{n,m}^{i(1)}$ | $h_{n,m}^{e(1)}$ and $\hat{h}_{n,m}^{e(1)}$ |
|---|---|---|---|
| $g_{1,0}^{i(1)}, g_{1,1}^{i(1)}, g_{2,0}^{i(1)}, g_{2,1}^{i(1)},$ | $g_{1,0}^{e(1)}, g_{1,1}^{e(1)}, g_{2,0}^{e(1)}, g_{2,1}^{e(1)},$ | $h_{2,0}^{i(1)}, h_{2,1}^{i(1)}, h_{2,2}^{i(1)}, h_{3,0}^{i(1)},$ | $h_{1,0}^{e(1)}, h_{1,1}^{e(1)}, h_{2,0}^{e(1)}, h_{2,1}^{e(1)},$ |
| $g_{2,2}^{i(1)}, g_{3,0}^{i(1)}, g_{3,1}^{i(1)}, g_{3,2}^{i(1)},$ | $g_{2,2}^{e(1)}, g_{3,0}^{e(1)}, g_{3,1}^{e(1)}, g_{3,2}^{e(1)},$ | $h_{3,1}^{i(1)}, h_{3,2}^{i(1)}, h_{3,3}^{i(1)}, \hat{h}_{2,1}^{i(1)},$ | $h_{2,2}^{e(1)}, h_{3,0}^{e(1)}, h_{3,1}^{e(1)}, h_{3,2}^{e(1)},$ |
| $g_{4,0}^{i(1)}, g_{4,1}^{i(1)}, g_{4,2}^{i(1)}, g_{4,3}^{i(1)},$ | $g_{4,0}^{e(1)}, g_{4,1}^{e(1)}, g_{4,2}^{e(1)}, g_{4,3}^{e(1)},$ | $\hat{h}_{2,2}^{i(1)}, \hat{h}_{3,1}^{i(1)}, \hat{h}_{3,2}^{i(1)}, \hat{h}_{3,3}^{i(1)}$ | $h_{3,3}^{e(1)}, \hat{h}_{1,1}^{e(1)}, \hat{h}_{2,1}^{e(1)}, \hat{h}_{2,2}^{e(1)},$ |
| $g_{4,4}^{i(1)}, \hat{g}_{1,1}^{i(1)}, \hat{g}_{2,1}^{i(1)}, \hat{g}_{2,2}^{i(1)},$ | $g_{4,4}^{e(1)}, \hat{g}_{1,1}^{e(1)}, \hat{g}_{2,1}^{e(1)}, \hat{g}_{2,2}^{e(1)},$ | | $\hat{h}_{3,1}^{e(1)}, \hat{h}_{3,2}^{e(1)}, \hat{h}_{3,3}^{e(1)}$ |
| $\hat{g}_{3,1}^{i(1)}, \hat{g}_{3,2}^{i(1)}, \hat{g}_{4,1}^{i(1)}, \hat{g}_{4,2}^{i(1)},$ | $\hat{g}_{3,1}^{e(1)}, \hat{g}_{3,2}^{e(1)}, \hat{g}_{4,1}^{e(1)}, \hat{g}_{4,2}^{e(1)},$ | | |
| $\hat{g}_{4,3}^{i(1)}, \hat{g}_{4,4}^{i(1)}$ | $\hat{g}_{4,3}^{e(1)}, \hat{g}_{4,4}^{e(1)}$ | | |

TABLE 2. Non-zero $g_{n,m}^{i(1)}, \hat{g}_{n,m}^{i(1)}, h_{n,m}^{i(1)}, \hat{h}_{n,m}^{i(1)}, g_{n,m}^{e(1)}, \hat{g}_{n,m}^{e(1)}, h_{n,m}^{e(1)}$ and $\hat{h}_{n,m}^{e(1)}$ obtained from the $O\left(Re_E\right)$ electric field.

$$g_{1,0}^{i(1)} = \frac{3S}{5\left(R+2\right)}\left(3E_x a_{2,1}^{(1)} + 3E_y \hat{a}_{2,1}^{(1)} + 2E_z a_{2,0}^{(1)}\right), \; g_{1,1}^{i(1)} = -\frac{3S}{5\left(R+2\right)}\left(E_x a_{2,0}^{(1)} - E_x a_{2,2}^{(1)} - 3E_z a_{2,1}^{(1)}\right),$$

$$\hat{g}_{1,1}^{i(1)} = -\frac{3S}{5\left(R+2\right)}\left(6E_y a_{2,2}^{(1)} - 3E_z \hat{a}_{2,1}^{(1)} + E_y a_{2,0}^{(1)}\right), \; g_{1,0}^{e(1)} = \frac{9\left(R+1\right)}{5\left(R+2\right)}\left(3E_y \hat{b}_{-3,1}^{(1)} + 2b_{3,0}^{(1)}E_z + 3E_x b_{-3,1}^{(1)}\right),$$

$$g_{1,1}^{e(1)} = \frac{9\left(R+1\right)}{5\left(R+2\right)}\left(3E_z b_{-3,1}^{(1)} + 6E_x b_{-3,2}^{(1)} - E_x b_{-3,0}^{(1)}\right), \; \hat{g}_{1,1}^{e(1)} = \frac{9\left(R+1\right)}{5\left(R+2\right)}\left(3E_z \hat{b}_{-3,1}^{(1)} - 6E_y b_{-3,2}^{(1)} - E_y b_{-3,0}^{(1)}\right).$$

$$\text{(F9)}$$

In a similar way as described for leading order approximation, using the boundary conditions (equation (2.12)) the coefficients of the solid spherical harmonics present in the Lamb solution of first order velocity and pressure field is obtained in terms of the unknown drop velocity as:

$$A_{n,m}^{(1)} = \frac{(2n+3)}{n(2n+1)(\lambda+1)}\Big[\big(4n^2+8n+3\big)\alpha_{n,m}^{(1)} + \big(4n^2-1\big)\beta_{n,m}^{(1)} + M\Big(g_{n,m}^{i(1)} - g_{n,m}^{e(1)}\Big)\Big],$$

$$A_{-n-1,m}^{(1)} = -\frac{1}{(n+1)(2n+1)(\lambda+1)}\Big[\lambda\big(8n^3+12n^2-2n-3\big)\alpha_{n,m}^{(1)} + (1+\lambda)\big(8n^2-2\big)\alpha_{-n-1,m}^{(1)}$$

$$+ \big\{\lambda\big(8n^2-2\big) + \big(8n^3+4n^2-2n-1\big)\big\}\beta_{n,m}^{(1)} + M\big(1-2n\big)\Big(g_{n,m}^{i(1)} - g_{n,m}^{e(1)}\Big)\Big],$$

$$B_{n,m}^{(1)} = -\frac{A_{n,m}^{(1)}}{2(2n+3)},$$

$$B_{-n-1,m}^{(1)} = -\frac{1}{2\lambda(2n^2+3n+1)}\Big[\big\{\lambda\big(4n^2+4n+1\big) - (4n+2)\big\}\alpha_{n,m}^{(1)} + \lambda\big(4n^2-1\big)\beta_{n,m}^{(1)}$$

$$- (1+\lambda)(4n+2)\beta_{-n-1,m}^{(1)} - M\Big(g_{n,m}^{i(1)} - g_{n,m}^{e(1)}\Big)\Big],$$

$$C_{n,m}^{(1)} = \frac{(2n+1)\gamma_{n,m}^{(1)} + M\Big(h_{n,m}^{e(1)} - h_{n,m}^{i(1)}\Big)}{n(n+1)(n+2+\lambda n-\lambda)},$$

$$C_{-n-1,m}^{(1)} = \frac{\Big[(n-1)(1-\lambda)\gamma_{n,m}^{(1)} + \big\{(2+n) + \lambda(n-1)\big\}\gamma_{-n-1,m}^{(1)} + M\Big(h_{n,m}^{e(1)} - h_{n,m}^{i(1)}\Big)\Big]}{n\big\{\big(n^2+3n+2\big) + \lambda\big(n^2-1\big)\big\}}, \tag{F8}$$

where $\beta_{1,0}^{(1)} = -U_{dz}^{(1)}, \beta_{1,1}^{(1)} = -U_{dx}^{(1)}$ and $\hat{\beta}_{1,1}^{(1)} = -U_{dy}^{(1)}$. The coefficients $\hat{A}_{n,m}^{(1)}, \hat{B}_{n,m}^{(1)}, \hat{C}_{n,m}^{(1)}, \hat{A}_{-n-1,m}^{(1)}$, $\hat{B}_{-n-1,m}^{(1)}$ and $\hat{C}_{-n-1,m}^{(1)}$ are obtained by substituting the terms $\beta_{n,m}^{(1)}, g_{n,m}^{i(1)}$ and $g_{n,m}^{e(1)}$ by $\hat{\beta}_{n,m}^{(1)}, \hat{g}_{n,m}^{i(1)}$ and $\hat{g}_{n,m}^{e(1)}$ respectively in equation (F8).

## REFERENCES

AHN, K., KERBAGE, C., HUNT, T.P., WESTERVELT, R.M., LINK, D.R. & WEITZ, D.A. 2006 Dielectrophoretic manipulation of drops for high-speed microfluidic sorting devices. *Applied Physics Letters* **88**(2), 024104.

BAROUD, C., DELVILLE, J.-P., GALLAIRE, F. & WUNENBURGER, R. 2007 Thermocapillary valve for droplet production and sorting. *Physical Review E* **75**(4), 046302.

BASU, A.S. & GIANCHANDANI, Y.B. 2008 Virtual microfluidic traps, filters, channels and pumps using Marangoni flows. *Journal of Micromechanics and Microengineering* **18**(11), 115031.

BHAGAT, A.A.S., BOW, H., HOU, H.W., TAN, S.J., HAN, J. & LIM, C.T. 2010 Microfluidics for cell separation. *Medical & biological engineering & computing* **48**(10), 999–1014.

BRINGER, M.R., GERDTS, C.J., SONG, H., TICE, J.D. & ISMAGILOV, R.F. 2004 Microfluidic systems for chemical kinetics that rely on chaotic mixing in droplets. *Philosophical* *transactions. Series A, Mathematical, physical, and engineering sciences* **362**(1818), 1087–104.

CASADEVALL I SOLVAS, X. & DEMELLO, A. 2011 Droplet microfluidics: recent developments and future applications. *Chemical communications (Cambridge, England)* **47**(7), 1936–42.

CHAN, P.C.-H. & LEAL, L.G. 2006 The motion of a deformable drop in a second-order fluid. *Journal of Fluid Mechanics* **92**(01), 131.

DESHMUKH, S.D. & THAOKAR, R.M. 2013 Deformation and breakup of a leaky dielectric drop in a quadrupole electric field. *Journal of Fluid Mechanics* **731**, 713–733.

FENG, J.Q. 1999 Electrohydrodynamic behaviour of a drop subjected to a steady uniform electric field at finite electric Reynolds number. *Proceedings of the Royal Society A: Mathematical, Physical and Engineering Sciences* **455**(1986), 2245–2269.

FENG, J.Q. & SCOTT, T.C. 2006 A computational analysis of electrohydrodynamics of a leaky dielectric drop in an electric field. *Journal of Fluid Mechanics* **311**(-1), 289.

FRANKE, T., ABATE, A.R., WEITZ, D.A. & WIXFORTH, A. 2009 Surface acoustic wave (SAW) directed droplet flow in microfluidics for PDMS devices. *Lab on a chip* **9**(18), 2625–7.

GRIGGS, A.J., ZINCHENKO, A.Z. & DAVIS, R.H. 2007 Low-Reynolds-number motion of a deformable drop between two parallel plane walls. *International Journal of Multiphase Flow* **33**(2), 182–206.

GUO, M.T., ROTEM, A., HEYMAN, J.A. & WEITZ, D.A. 2012 Droplet microfluidics for high-throughput biological assays. *Lab on a chip* **12**(12), 2146–55.

HABER, S. & HETSRONI, G. 1971 The dynamics of a deformable drop suspended in an unbounded Stokes flow. *Journal of Fluid Mechanics* **49**(02), 257–277.

HANNA, J.A. & VLAHOVSKA, P.M. 2010 Surfactant-induced migration of a spherical drop in Stokes flow. *Physics of Fluids* **22**(1), 013102.

HAPPEL, J. & BRENNER, H. 1981 *Low Reynolds number hydrodynamics*, Dordrecht: Springer Netherlands.

HE, H., SALIPANTE, P.F. & VLAHOVSKA, P.M. 2013 Electrorotation of a viscous droplet in a uniform direct current electric field. *Physics of Fluids* **25**(3), 032106.

HETSRONI, G. & HABER, S. 1970 The flow in and around a droplet or bubble submerged in an unbound arbitrary velocity field. *Rheologica Acta* **9**(4), 488–496.

HETSRONI, G., HABER, S. & WACHOLDER, E. 1970 The flow fields in and around a droplet moving axially within a tube. *Journal of Fluid Mechanics* **41**(04), 689–705.

KIM, S. & KARRILA, S. 1991 *Microhydrodynamics : Principles and Selected Applications*, London: Butterworth-Heinemann.

LAC, E. & HOMSY, G.M. 2007 Axisymmetric deformation and stability of a viscous drop in a steady electric field. *Journal of Fluid Mechanics* **590**.

LAMB, H. 1975 *Hydrodynamics* 6th ed., London: Cambridge University Press.

LANAUZE, J. A., WALKER, L.M. & KHAIR, A.S. 2013 The influence of inertia and charge relaxation on electrohydrodynamic drop deformation. *Physics of Fluids* **25**(11), 112101.

LEAL, L.G. 2007 *Advanced Transport Phenomena*, Cambridge: Cambridge University Press.

LINK, D.R., GRASLAND-MONGRAIN, E., DURI, A., SARRAZIN, F., CHENG, Z., CRISTOBAL, G., MARQUEZ, M. & WEITZ, D.A. 2006 Electric control of droplets in microfluidic devices. *Angewandte Chemie (International ed. in English)* **45**(16), 2556–60.

MAGNAUDET, J. 2003 Small inertial effects on a spherical bubble, drop or particle moving near a wall in a time-dependent linear flow. *Journal of Fluid Mechanics* **485**, 115–142.

MELCHER, J.R. & TAYLOR, G.I. 1969 Electrohydrodynamics: A Review of the Role of Interfacial Shear Stresses. *Annual Review of Fluid Mechanics* **1**(1), 111–146.

MORTAZAVI, S. & TRYGGVASON, G. 2000 A numerical study of the motion of drops in Poiseuille flow. Part 1. Lateral migration of one drop. *Journal of Fluid Mechanics* **411**, 325–350.

MUKHERJEE, S. & SARKAR, K. 2013 Effects of matrix viscoelasticity on the lateral migration of a deformable drop in a wall-bounded shear. *Journal of Fluid Mechanics* **727**, 318–345.

MUKHERJEE, S. & SARKAR, K. 2014 Lateral migration of a viscoelastic drop in a Newtonian fluid in a shear flow near a wall. *Physics of Fluids* **26**(10), 103102.

PAK, O.S., FENG, J. & STONE, H.A. 2014 Viscous Marangoni migration of a drop in a Poiseuille flow at low surface Péclet numbers. *Journal of Fluid Mechanics* **753**, 535–552.

PAMME, N. 2012 On-chip bioanalysis with magnetic particles. *Current opinion in chemical biology* **16**(3-4), 436–43.

SALIPANTE, P.F. & VLAHOVSKA, P.M. 2010 Electrohydrodynamics of drops in strong uniform dc electric fields. *Physics of Fluids* **22**(11), 112110.

SAVILLE, D.A. 1997 ELECTROHYDRODYNAMICS:The Taylor-Melcher Leaky Dielectric Model. *Annual Review of Fluid Mechanics* **29**(1), 27–64.

SCHWALBE, J.T., VLAHOVSKA, P.M. & MIKSIS, M.J. 2011 Vesicle electrohydrodynamics. *Physical Review E* **83**(4), 046309.

SEEMANN, R., BRINKMANN, M., PFOHL, T. & HERMINGHAUS, S. 2012 Droplet based microfluidics. *Reports on progress in physics* **75**(1), 016601.

SUBRAMANIAN, R.S. & BALASUBRAMANIAM, R. 2005 *The Motion of Bubbles and Drops in Reduced Gravity*, Cambridge: Cambridge University Press.

SUPEENE, G., KOCH, C.R. & BHATTACHARJEE, S. 2008 Deformation of a droplet in an electric field: nonlinear transient response in perfect and leaky dielectric media. *Journal of colloid and interface science* **318**(2), 463–76.

TAYLOR, G. 1966 Studies in Electrohydrodynamics. I. The Circulation Produced in a Drop by Electrical Field. *Proceedings of the Royal Society A: Mathematical, Physical and Engineering Sciences* **291**(1425), 159–166.

TEH, S.-Y., LIN, R., HUNG, L.-H. & LEE, A.P. 2008 Droplet microfluidics. *Lab on a chip* **8**(2), 198–220.

TORZA, S., COX, R.G. & MASON, S.G. 1971 Electrohydrodynamic Deformation and Burst of Liquid Drops. *Philosophical Transactions of the Royal Society A: Mathematical, Physical and Engineering Sciences* **269**(1198), 295–319.

VIZIKA, O. & SAVILLE, D. A. 2006 The electrohydrodynamic deformation of drops suspended in liquids in steady and oscillatory electric fields. *Journal of Fluid Mechanics* **239**(-1), 1.

VLAHOVSKA, P.M. 2011 On the rheology of a dilute emulsion in a uniform electric field. *Journal of Fluid Mechanics* **670**, 481–503.

VLAHOVSKA, P.M., LOEWENBERG, M. & BLAWZDZIEWICZ, J. 2005 Deformation of a surfactant-covered drop in a linear flow. *Physics of Fluids* **17**(10), 103103.

WARD, T. & HOMSY, G.M. 2006 Chaotic streamlines in a translating drop with a uniform electric field. *Journal of Fluid Mechanics* **547**(-1), 215.

WARD, T. & HOMSY, G.M. 2001 Electrohydrodynamically driven chaotic mixing in a translating drop. *Physics of Fluids* **13**(12), 3521.

WOHL, P.R. & RUBINOW, S.I. 2006 The transverse force on a drop in an unbounded parabolic flow. *Journal of Fluid Mechanics* **62**(01), 185.

XU, X. & HOMSY, G.M. 2006 The settling velocity and shape distortion of drops in a uniform electric field. *Journal of Fluid Mechanics* **564**, 395.

XU, X. & HOMSY, G.M. 2007 Three-dimensional chaotic mixing inside drops driven by a transient electric field. *Physics of Fluids* **19**(1), 013102.

ZHENG, B., TICE, J.D. & ISMAGILOV, R.F. 2004 Formation of droplets of alternating composition in microfluidic channels and applications to indexing of concentrations in droplet-based assays. *Analytical chemistry* **76**(17), 4977–82.

ZHU, Y. & FANG, Q. 2013 Analytical detection techniques for droplet microfluidics--a review. *Analytica chimica acta* **787**, 24–35.

ZHU, Y., ZHU, L.-N., GUO, R., CUI, H.-J., YE, S. & FANG, Q. 2014 Nanoliter-scale protein crystallization and screening with a microfluidic droplet robot. *Scientific reports* **4**, 5046.